\documentclass[a4paper,11pt]{article}
\pdfoutput=1

\usepackage{jheppub}
\usepackage{pdflscape}
\usepackage{color}
\usepackage{ulem}
\usepackage{tabularx}
\usepackage{hyperref}
\usepackage{amsmath}
\usepackage{amssymb}
\usepackage{graphicx,multirow,subfigure}
\usepackage[T1]{fontenc}
\usepackage{comment}
\usepackage[dvipsnames]{xcolor}
\usepackage{physics}
\usepackage{bbold}
\usepackage{mathtools}
\usepackage{cancel}

\renewcommand{\bar}{\overline}

\renewcommand \ket[1]{
        \left| #1 \right>
}

\renewcommand \bra[1]{
        \left< #1 \right|
}

\newcommand{\bi}{\begin{itemize}}
\newcommand{\ei}{\end{itemize}}
\newcommand{\ben}{\begin{enumerate}}
\newcommand{\een}{\end{enumerate}}
\newcounter{mycount}
\newcommand{\pauseen}{\setcounter{mycount}{\value{enumi}}\end{enumerate}}
\newcommand{\resumeen}{\begin{enumerate}\setcounter{enumi}{\value{mycount}}}

\newcommand{\be}{\begin{equation}}
\newcommand{\ee}{\end{equation}}
\newcommand{\bea}{\begin{eqnarray}}
\newcommand{\eea}{\end{eqnarray}}
\newcommand{\beq}{\begin{equation}}
\newcommand{\eeq}{\end{equation}}
\def\beqa{\begin{eqnarray}}
  \def\eeqa{\end{eqnarray}}
\def\lsim{\mathrel{\rlap{\lower4pt\hbox{\hskip1pt$\sim$}}
    \raise1pt\hbox{$<$}}}         
\def\gsim{\mathrel{\rlap{\lower4pt\hbox{\hskip1pt$\sim$}}
    \raise1pt\hbox{$>$}}}         

\begin{document}

\title{CP Violation in Charmed Meson Decays into Final States with $\eta'$}

\author[a,b]{Carolina Bolognani}
\author[c]{Ulrich Nierste}
\author[d]{Stefan Schacht}
\author[a,b]{K. Keri Vos}

\affiliation[a]{Gravitational Waves and Fundamental Physics (GWFP), Maastricht University, Duboisdomein 30, NL-6229 GT Maastricht, the Netherlands}
\affiliation[b]{Nikhef, Science Park 105, NL-1098 XG Amsterdam, the Netherlands}
\affiliation[c]{Institute for Theoretical Particle Physics, Karlsruhe Institute of Technology (KIT), Wolfgang-Gaede-Str. 1, 76131 Karlsruhe, Germany}
\affiliation[d]{Institute for Particle Physics Phenomenology, Department of Physics, Durham University, Durham DH1 3LE, United Kingdom}

\emailAdd{carolina.bolognani@cern.ch}
\emailAdd{ulrich.nierste@kit.edu}
\emailAdd{stefan.schacht@durham.ac.uk}
\emailAdd{k.vos@maastrichtuniversity.nl}

\subheader{\hfill \textnormal{Nikhef 2025-018, TTP25-044, IPPP/25/73}}

\abstract{
We derive Standard Model predictions for the CP asymmetries of singly-Cabibbo suppressed $D\rightarrow P\eta'$ decays, where $P=K,\pi,\eta$. Our predictions are based on the approximate SU(3)$_F$ symmetry of QCD and include first-order symmetry-breaking effects in a systematic way.
The underlying symmetry leads to correlations between different decay modes.
To this end we predict the correlations between $ a_{\rm CP}^{\rm dir}(D^+\to \pi^+ \eta')$ and  $a_{\rm CP}^{\rm dir}(D_s^+ \to K^+ \eta')$ as well as  $a_{\rm CP}^{\rm dir}(D^0\to \pi^0 \eta')$ and $ a_{\rm CP}^{\rm dir}(D^0\to \eta \eta')$. Our results can be used to probe the Standard Model with future measurements by LHCb, Belle II and BESIII. At the same time, such future measurements permit the extraction of the key theory parameter related to the ratio of color-suppressed to color-favoured contributions to the decay amplitudes. Future data on both branching ratios and CP asymmetries will automatically improve our predictions and thereby also their sensitivity to physics beyond the Standard Model.
}

\maketitle
\flushbottom

\section{Introduction \label{sec:introduction}}

\normalem

Charm CP violation is a very active field, with many on-going searches~\cite{LHCb:2023qne,LHCb:2023mwc, LHCb:2023rae, LHCb:2024rkp,CMS:2024hsv, Belle-II:2023vra, Belle:2023bzn,LHCb:2025ezf} that follow the discovery of charm CP violation~\cite{LHCb:2019hro} and the subsequent measurement of  
unexpectedly large CP violation in the $D^0\rightarrow \pi^+\pi^-$ decay channel~\cite{LHCb:2022lry}. 
Recent measurements also include decays to final states with $\eta^{(\prime)}$~\cite{BESIII:2022xhe, Belle:2021dfa, LHCb:2022pxf, LHCb:2021rou, Belle:2021ygw} and are thus related to the subject of this paper.
Future prospects for such decays have been investigated as well~\cite{Belle-II:2018jsg, Cerri:2018ypt, LHCb:2018roe}.

CP asymmetries are of particular theoretical importance as they permit the extraction and interpretation of the ratio of CKM-subleading (penguin) contributions over CKM-leading (tree) ones, see Refs.~\cite{Brod:2011re, Brod:2012ud, Franco:2012ck, Cheng:2012wr, Khodjamirian:2017zdu, Cheng:2019ggx, Chala:2019fdb, Grossman:2019xcj, Dery:2021mll, Schacht:2021jaz, Schacht:2022kuj, Bediaga:2022sxw, Lenz:2023rlq, Gavrilova:2023fzy, Pich:2023kim} for recent studies. Interpretations in terms of 
new physics parameters have also been discussed~\cite{Grossman:2006jg, Altmannshofer:2012ur, Bause:2022jes, Dery:2019ysp}. 
A very useful tool in this context are sum rules between 
different channels~\cite{Grossman:2012eb, Grossman:2012ry, Grossman:2013lya, Muller:2015rna, Grossman:2018ptn, Gavrilova:2022hbx, Gavrilova:2024npn, Iguro:2024uuw} 
which are based on SU(3)$_F$ symmetry and its subgroups. 
The latter also allows global fits of the group theoretical and topological parametrizations~\cite{Pirtskhalava:2011va, Feldmann:2012js, Bhattacharya:2012ah, Hiller:2012xm, Muller:2015lua, Muller:2015rna,  Nierste:2015zra}. A direct calculation of the penguin over tree ratio from first principles may one day be possible using Lattice QCD~\cite{Hansen:2012tf, DiCarlo:2025mnm}. 

We analyzed the branching ratios of $D\to P\eta'$, where $P=K,\pi, \eta$, using $SU(3)_F$ symmetry including linear $SU(3)_F$-breaking corrections in our previous work~\cite{Bolognani:2024zno}. In this work, we extend this analysis and focus on predictions for CP asymmetries in decays with $\eta'$ final states.
Since $\eta'$ is a flavour singlet in the limit of exact SU(3)$_F$ symmetry, the symmetry structure of the decay amplitudes is particularly simple and implies relations (sum rules) 
between different CP asymmetries, which can be tested to discover or constrain new physics. Clearly, 
robust predictions for these sum rules need reliable constraints on the corrections from 
SU(3)$_F$ breaking, which we explore in this paper. The presence of  $\eta$-$\eta^{\prime}$ mixing 
poses a challenge, because the largely different $\eta$ and $\eta'$ masses obscure connections of $D$ decay matrix elements with $\eta$ to those with $\eta^\prime$.
The formalism for the treatment of $D$ decays to $\eta^{(\prime)}$ final states has been laid out in our previous work~\cite{Bolognani:2024zno}.
As described in detail in Ref.~\cite{Bolognani:2024zno}, in our approach we aim to use only the SU(3)$_F$ expansion with, as much as possible, no additional theory input beyond that, leading to a complete decoupling of the $D\rightarrow P\eta'$ system from $D\rightarrow P\eta$ and $D\rightarrow PP$. In this way, our approach differs from previous studies~\cite{Bhattacharya:2009ps, Grossman:2012ry, Bhattacharya:2021ndt}.

The CP asymmetries of $D\rightarrow P\eta'$ are, with the exception 
of one measurement, basically uncharted territory and open a lot of opportunities for 
future measurements that can be compared to our predictions. Our results can be used in order to extract the ratio of CKM-subleading over CKM-leading amplitude ratios from future data and probe the consistency of the data.  

In Sec.~\ref{sec:notation} we introduce our notation. 
In Sec.~\ref{sec:decomposition} we present our results for the topological decomposition of the CKM-subleading amplitudes.
In particular we show how to account for the linear dependences of the unknown hadronic parameters through redefinitions which combine several parameters into a single one. 
In Sec.~\ref{sec:predictions}, we present our predictions for the CP asymmetries, which can be used in the future to probe the SM and learn more about the quality of the $1/N_c$ expansion in QCD,
where $N_c=3$ denotes the number of colours. We conclude in Sec.~\ref{sec:conclusions}. In the appendix, we give the full parametrization of the CKM-subleading amplitudes without redefinitions.

\section{\boldmath Notation \label{sec:notation}}

We use the notation introduced in Ref.~\cite{Bolognani:2024zno} and recall here only some key elements.
The underlying quark-level transitions of singly Cabibbo-suppressed (SCS) charm decays are $c\to d\bar{d}u$ and $c\to s \bar{s}u$. 
We write the amplitude of SCS decays as
\begin{align}
\mathcal{A}^{\mathrm{SCS}}(d) &= \lambda_{sd} \mathcal{A}_{sd}(d) - \frac{\lambda_b}{2} \mathcal{A}_b(d)\,, \label{eq:SCSdecays}
\end{align}
with
\begin{align}
\lambda_{sd} &\equiv \frac{\lambda_s - \lambda_d}{2}\,, \\
-\frac{\lambda_b}{2} &= \frac{\lambda_s+\lambda_d}{2}\,,
\end{align}
and the combinations of elements of the Cabibbo-Kobayashi-Maskawa (CKM) matrix
\begin{align}
\lambda_q &\equiv V_{cq}^*V_{uq}\,.
\end{align}
Consequently, we write the CKM factors that appear in $c\to s\bar{s}u$ and $c\to d \bar{d}u$ transitions as
\begin{align}
    \lambda_s = \lambda_{sd}-\frac{\lambda_b}{2},\\
    \lambda_d = -\lambda_{sd}-\frac{\lambda_b}{2},
\end{align}
respectively. The Cabibbo suppression of $\lambda_b$ versus $\lambda_{sd}$ makes the $\mathcal{A}_b(d)$ contribution to the branching ratios negligible, \emph{i.e.}, we have to a very good approximation 
\begin{align}
\mathcal{B}(d) &= \mathcal{P}(d)\left|\lambda_{sd}\mathcal{A}_{sd}(d)\right|^2\,, \label{eq:normalization}
\end{align}
where the phase space function is given as~\cite{ParticleDataGroup:2024cfk,Muller:2015lua}
\begin{align}
\mathcal{P}(d) &= \frac{\tau_D}{16 \pi m_D^3} \sqrt{
		\left(m_D^2 - ( m_{P_1} - m_{P_2} )^2 \right)
		\left(m_D^2 - ( m_{P_1} + m_{P_2} )^2 \right)
			}\,. \label{eq:phase-space}
\end{align}
Eq.~(\ref{eq:SCSdecays}) makes clear that in SCS charm decays, CP violation emerges from the interference of CKM-leading with the CKM-subleading amplitudes.
The direct CP asymmetries are defined as
\begin{equation}
    a_{CP}^\textrm{dir}(d) = \frac{\left|\mathcal{A}^\text{SCS}(d)\right|^2 - \left|\mathcal{\bar{A}}^\text{SCS}(\bar{d})\right|^2}{\left|\mathcal{A}^\text{SCS}(d)\right|^2 + \left|\mathcal{\bar{A}}^\text{SCS}(\bar{d})\right|^2},
\end{equation}
and, neglecting subleading powers of $\lambda_b/\lambda_{sd}\sim 10^{-3}$, given as~\cite{Golden:1989qx, Pirtskhalava:2011va, Nierste:2017cua} 
\begin{equation}
	a_{CP}^\textrm{dir}(d) = \mathrm{Im}\frac{\lambda_b}{\lambda_{sd}} \mathrm{Im}\frac{\mathcal{A}_b(d)}{\mathcal{A}_{sd}(d)} \label{eq:aCPdir}
\end{equation}
which depends on the relative strong phase between the CKM-leading and -subleading amplitude.
For the $\ket{\eta_8}$ and $\ket{\eta_0}$ states we use the definitions 
\begin{align}
\ket{\eta_8} &\simeq \frac{\ket{u\bar u}+ \ket{d\bar d} -2 \ket{s\bar s}}{\sqrt6}\,, \\
\ket{\eta_0} &\simeq \frac{\ket{u\bar{u}}+\ket{d\bar{d}}+\ket{s \bar{s}}}{\sqrt{3}}\,,
\end{align}
respectively. Furthermore, we write the hadronic matrix elements as 
\begin{align}
  \bra{P\eta}H\ket{D}
  &=\, \cos\theta \bra{P\eta_8} H \ket{D}  -
        \sin\theta \bra{P\eta_0} H \ket{D}     \label{eq:eta-mixing-1}\\
  \bra{P\eta'}H\ket{D}
  &=\,  \sin\theta\bra{P\eta_8} H\ket{D}'+
    \cos\theta \bra{P\eta_0} H\ket{D}'\,, \label{eq:eta-mixing-2}
\end{align}
with a mixing angle that respects the SU(3)$_F$ power counting 
$\theta = \mathcal{O}(\varepsilon)$, where $\varepsilon\sim 30\%$ is the parameter of SU(3)$_F$ breaking. The mixing angle $\theta$ will subsequently always appear together with
SU(3)$_F$-breaking matrix elements into which it can be absorbed.
As laid out in detail in Ref.~\cite{Bolognani:2024zno}, the large mass splitting  $m_{\eta}\neq m_{\eta^\prime}$ leads to the inequalities
\begin{align}
\bra{P\eta_8} H \ket{D}' \neq \bra{P\eta_8} H\ket{D}\,,\\
\bra{P\eta_0} H\ket{D}' \neq \bra{P\eta_0} H\ket{D}\,.
\end{align}
The difference between these matrix elements is an $\mathcal{O}(1)$ effect that goes beyond the SU(3)$_F$ power counting, \emph{i.e.},~the LHS and RHS will not become equal in the limit $\varepsilon\to 0$.
This leads to a decoupling of the $D\rightarrow P\eta'$ system from $D\rightarrow P\eta$ and $D\rightarrow PP$, 
as long as no further assumptions on the dependences of the matrix elements on $m_{\eta^{(\prime)}}$ are adopted. 

Potential new-physics contributions could change the relations between the decay amplitudes that hold in the SM. The SM CP asymmetries stem from the interference of 
the tiny second term $\propto \lambda_b \mathcal{A}_b$ with the dominant tree-level amplitude $\propto \lambda_{sd} \mathcal{A}_{sd}$. A new-physics contribution may numerically compete with 
the $\lambda_b \mathcal{A}_b$ term and is indistinguishable from the SM contribution as long as one only considers a single CP asymmetry. SU(3)$_F$ symmetry, however, correlates the topological amplitudes of the different decays with each other. For example,  $ \mathcal{A}_b(d)$ involves the piece of the effective SM decay Hamiltonian which is a singlet of the U-spin subgroup of SU(3)$_F$. A new-physics contribution with a U-spin non-singlet interaction (and a CP phase different from $\arg(\pm \lambda_{sd})$) will violate the amplitude relations which give rise to the SM predictions presented in this paper (see Ref.~\cite{Iguro:2024uuw} for a study without $\eta^\prime$ final states). One can see this in an easy way from $a_{\rm CP}(D^0\to K^+K^-)$ and $a_{\rm CP}(D^0\to \pi^+\pi^-)$, where a potentially enhanced SM contribution  always contributes with opposite signs to these CP asymmetries, while $\Delta U=1$  new physics contributes with the same sign, see Refs.~\cite{Hiller:2012xm,  Iguro:2024uuw, Bause:2022jes, Schacht:2022kuj}.

\section{Topological Decomposition of CKM-subleading Amplitudes \label{sec:decomposition} }

\subsection{Decomposition}

The topological decomposition of the CKM-leading amplitudes $\mathcal{A}_{sd}(d)$ is given in Ref.~\cite{Bolognani:2024zno}, parametrized by topological tree (T), colour-suppressed (C), annihilation (A), and exchange (E) topologies. Here, we derive the corresponding CKM-subleading amplitudes $\mathcal{A}_b(d)$, which, through interference with the leading amplitudes lead to CP violation.
We show our result for the complete decomposition of $\mathcal{A}_b(d)$ in terms of topological amplitudes $\mathcal{T}_i$ and the $SU(3)_F$-breaking topologies $\mathcal{T}_i^{(1)}$ in 
Tables~\ref{tab:Ab-Ts-As}, \ref{tab:Ab-Cs-Es} and \ref{tab:Ab-Ps} in Appendix~\ref{sec:tables}. We note that the subleading amplitude $\mathcal{A}_b$ consists of a combination of topologies already present in $\mathcal{A}_{sd}(d)$ and defined in Ref.~\cite{Bolognani:2024zno} and additional topologies that only contribute to $\mathcal{A}_b(d)$ that we define in Fig.~\ref{fig:Absuppressedtopologies}. We label the topologies depending on whether the $\eta_0$ singlet is formed by the outgoing quark or antiquark by $18$ and $81$, respectively. In addition, the $11$ and $88$ subscript indicate the contribution from the suppressed singlet state of the $\eta$ and the suppressed octet state of the $\eta'$, respectively. Notably, additional topologies that appear are penguin (P) and penguin annihilation (PA) diagrams, for which we use the notation~\cite{Muller:2015rna}
\begin{align}
P	   &\equiv P_d+P_s-2P_b\,,\\
PA	   &\equiv PA_d+PA_s-2PA_b\,,
\end{align}
where the respective subscript indicates the flavour of the quark running in the loop. Analogous definitions apply to $P_{88}^{(1)}$, $P_{11}^{(1)}$ and $P_H$. Here, $P_H$ are the hairpin penguin diagrams, \emph{i.e.},~penguin diagrams with a disconnected part, which do not appear in the $D\rightarrow PP$ system. 
Hairpin diagrams have been discussed in the literature for a long time, see \emph{e.g.}~Ref.~\cite{Lindenbaum:1980rf} for a discussion in the context of the OZI rule in $\phi$ decays and Refs.~\cite{Du:1993qe, Buras:1998ra} for the role of hairpin diagrams in the context of non-leptonic $B$ decays.
Like penguin diagrams, such diagrams also contribute to the branching ratios, but only in the combination
\begin{align}
P_{H, \mathrm{break}} \equiv P_{H,s}-P_{H,d}\,,
\end{align}
where the index denotes the quark that runs in the loop.
The broken hairpin penguin diagrams contributing to the CKM-leading amplitudes
\begin{align}
\mathcal{A}_{sd}(D^0\rightarrow \pi^0 \eta') &\supset \frac{3}{\sqrt{6}} P_{H,\mathrm{break}}^{(1)}\,,\\
\mathcal{A}_{sd}(D^0\rightarrow \eta\eta')   &\supset \frac{1}{\sqrt{2}} P_{H,\mathrm{break}}^{(1)}\,,\\
\mathcal{A}_{sd}(D^+\rightarrow \pi^+\eta')  &\supset \frac{3}{\sqrt{3}} P_{H,\mathrm{break}}^{(1)}\,, \\
\mathcal{A}_{sd}(D_s^+\rightarrow K^+\eta )  &\supset \frac{3}{\sqrt{3}} P_{H,\mathrm{break}}^{(1)}\,,
\end{align}
can be absorbed into $P^{(1)}_{18,\mathrm{break}}$ and $P^{(1)}_{81,\mathrm{break}}$, as done implicitly in Ref.~\cite{Bolognani:2024zno}.  In the following, we make the dependence on the broken hairpin penguin explicit.

For the direct CP asymmetries the crucial aspect is the interference between $\mathcal{A}_b(d)$ and $\mathcal{A}_{sd}(d)$.
Therefore, it is convenient to write $\mathcal{A}_b(d)$ in terms of $\mathcal{A}_{sd}(d)$, plus a remainder that we call $\Delta \mathcal{A}_b(d)$~\cite{Muller:2015rna}
\begin{align}
     \mathcal{A}_b(d)&=c_{sd}^d \mathcal{A}_{sd}(d) + \Delta \mathcal{A}_b(d)\\
     & = c_{sd}^d \mathcal{A}_{sd}(d) + \sum_i c_i^d\mathcal{T}_i,\label{eq:deltaAb}
\end{align}
where $c_{sd}^d$ are coefficients that depend on the decay $d$ that can be inferred from Tables~\ref{tab:Ab-Ts-As}, \ref{tab:Ab-Cs-Es} and \ref{tab:Ab-Ps} in Appendix~\ref{sec:tables}. Due to linear dependences of the parameters in $\Delta \mathcal{A}_b$, we can reduce the number of parameters therein by applying redefinitions.
For consistency, we apply the same redefinitions as given in Eqs.~(3.4), (3.5), and (3.9)--(3.15) in Ref.~\cite{Bolognani:2024zno}. 
For completeness, we quote the redefinitions relevant for the parameters that enter $\Delta \mathcal{A}_b(d)$, 
\begin{align}
        \hat{C}_{81} & = C_{81}+E_{18}+E_{81}+3E_H\,,\label{eq:oldredefs-1}\\
        \hat{C}_{88}^{(1)} &=C_{88}^{(1)} - E_{88}^{(1)}\,,\label{eq:oldredefs-2} \\
        \hat{C}_{18,1}^{(1)}&=C_{18,1}^{(1)} + C_{18,2}^{(1)}+P_{18,\mathrm{break}}^{(1)} + P_{81,\mathrm{break}}^{(1)} + 3 P_{H,\mathrm{break}}^{(1)}\,. \label{eq:oldredefs-3}
\end{align}
As mentioned above, we now show the broken hairpin penguin $P_{H,\mathrm{break}}^{(1)}$ explicitly.

The additional diagrams appearing in the CKM-subleading amplitudes make additional redefinitions necessary. 
Preserving the power-counting by not mixing $SU(3)_F$ limit parameters with $SU(3)_F$-breaking ones, we use additional redefinitions. For the $SU(3)_F$ limit parameter, we redefine
\begin{align}
        \bar{C}_{18} & = C_{18} + \frac{P_{18}+P_{81}+3P_H}{2}\,, \label{eq:redef-penguin}
        \end{align}
which reabsorbs all the penguin parameters into $\bar{C}_{18}$. This also makes it impossible to distinguish between penguins and colour-suppressed topologies. 

For the $SU(3)_F$-breaking parameters, we redefine        
    \begin{align}
        \bar{C}_{18,3}^{(1)}  & = C_{18,3}^{(1)} + \frac{P_{18,2}^{(1)}+P_{81,1}^{(1)}+P_{81,2}^{(1)}+3P_{H,1}^{(1)}}{2} \nonumber \\ 
        & \quad -\frac{3}{4}\left( P_{18,\mathrm{break}}^{(1)} + P_{81,\mathrm{break}}^{(1)}+3P^{(1)}_{H,\mathrm{break}} \right)\,,\\
        \bar{C}_{11}^{(1)}  & = C_{11}^{(1)} + E_{11}^{(1)}+3E_{H,11}^{(1)}+\frac{P_{11}^{(1)}+3P_{H,11}^{(1)}+3PA_{11}^{(1)}+9PA_{H,11}^{(1)}}{2} \nonumber \\
        &  \quad - \frac{\sqrt{2}}{4}\left( PA_{18,1}^{(1)}+PA_{18,2}^{(1)}+PA_{81,1}^{(1)}+PA_{81,2}^{(1)}+3PA_{H,1}^{(1)}\right) \nonumber \\
        & \quad  -\frac{3}{4\sqrt{2}}\left( P_{18,\mathrm{break}}^{(1)} + P_{81,\mathrm{break}}^{(1)}+3P^{(1)}_{H,\mathrm{break}} \right) + \frac{3}{2}PA_{88}^{(1)} + \frac{3}{4}P_{88}^{(1)}\,,\\
        \bar{E}_{88}^{(1)} & = E_{88}^{(1)} - \frac{P_{88}^{(1)}}{2}+\frac{1}{2\sqrt{2}}\left( P_{18,\mathrm{break}}^{(1)} + P_{81,\mathrm{break}}^{(1)}+3P^{(1)}_{H,\mathrm{break}} \right)\,.
        \label{eq:newredefs2ndTry}
\end{align}

The result of all redefinitions for the coefficients ${c}_{sd}^d$ and ${c}^d_i$ given in Eq.~\eqref{eq:deltaAb} is shown in Table~\ref{tab:redefparams}. 
Further redefinitions could still be made between $SU(3)_F$-breaking parameters, which would imply absorbing $\hat{C}_{18,1}^{(1)}$ and $\hat{C}_{88}^{(1)}$ into new parameters. We chose, however, to keep their explicit dependence on the parametrisation as these parameters have additional constraints from the leading $\mathcal{A}_{sd}$ amplitude fit, in Ref.~\cite{Bolognani:2024zno}.
We note that all redefined SU(3)$_F$ limit topologies also include hairpin diagrams. We do not impose additional theoretical constraints on these diagrams, like for example OZI suppression~\cite{Okubo:1963fa, Iizuka:1966fk, Zweig:1964jf}. Due to the linear dependences it is not feasible to study the size of the hairpin diagrams separately from the other diagrams without invoking further theoretical assumptions that would relate the $D\rightarrow P\eta'$ system to $D\rightarrow P\eta$ and $D\rightarrow PP$ decays.

\begin{table}[t]
\begin{center}
\setlength{\tabcolsep}{0.5pt} 
\begin{tabular}{c|c|cc|ccccc}\hline\hline
 Decay ampl. {$\mathcal{A}_b(d)$}  & $\mathcal{A}_{sd}$ & $\bar{C}_{18}$ & $\hat{C}_{81}$ & $\hat{C}_{18,1}^{(1)}$ & $\bar{C}_{18,3}^{(1)}$ & $\hat{C}_{88}^{(1)}$ &$\bar{C}_{11}^{(1)}$ & $\bar{E}_{88}^{(1)}$\\ \hline\hline
 $D^0 \rightarrow \pi^0 \eta' $ & $1$ & $\sqrt{\frac{2}{3}}$   &$-\sqrt{\frac{2}{3}}$ & 0 & 0 & 0 & 0 & $-\frac{2}{\sqrt{3}}$ 
 \\
$D^0 \rightarrow \eta \eta' $ & $1$ & $\frac{\sqrt{2}}{3}$  &$\frac{\sqrt{2}}{3}$  & 0 & 0 &   $\frac{2}{3}$ & $\frac{4}{3}$  & $\frac{4}{3}$   \\

$D^+ \rightarrow \pi^+ \eta' $ & $-1$ &  $\frac{2}{\sqrt{3}}$ &
0 &  
$\frac{2}{\sqrt{3}}$ & 0 & 
$-2\sqrt{\frac{2}{3}}$ & 0 & $-2\sqrt{\frac{2}{3}}$  \\

 $D_s^+ \rightarrow K^+ \eta' $ &  $1$ & $\frac{2}{\sqrt{3}}$ & 0 & 0 & $\frac{2}{\sqrt{3}}$ & $\sqrt{\frac{2}{3}}$ & 0 & $\sqrt{\frac{2}{3}}$ \\\hline\hline
    \end{tabular}
\caption{Coefficients of the topological decomposition of the CKM-subleading amplitudes $\mathcal{A}_{b}(d)$ in Eq.~(\ref{eq:deltaAb}).
} 
\label{tab:redefparams}
\end{center}
\end{table}

\subsection{CP asymmetry sum rules}

From the redefined topological decomposition given in Table~\ref{tab:redefparams} as well as the corresponding table for the CKM-leading amplitudes in Ref.~\cite{Bolognani:2024zno}, we can also read off the $SU(3)_F$ limit amplitudes
\begin{align}
        \mathcal{A}^{\text{SCS}}_{\text{SU(3)$_F$-lim.}}(D^0\to \pi^0\eta') & = \lambda_{sd}\left(\frac{1}{\sqrt{6}}\hat{C}_{81} \right) -\frac{\lambda_b}{2}\left(\frac{2}{\sqrt{6}}\bar{C}_{18} - \frac{1}{\sqrt{6}}\hat{C}_{81} \right)\,, \label{eq:su3-limit-amp-1}\\
        \mathcal{A}^{\text{SCS}}_{\text{SU(3)$_F$-lim.}}(D^0\to \eta\eta') & = \lambda_{sd}\left(-\frac{1}{\sqrt{2}}\hat{C}_{81} \right) -\frac{\lambda_b}{2}\left(\frac{2}{3\sqrt{2}}\bar{C}_{18} - \frac{1}{3\sqrt{2}}\hat{C}_{81} \right)\,, \label{eq:su3-limit-amp-2}\\
        \mathcal{A}^{\text{SCS}}_{\text{SU(3)$_F$-lim.}}(D^+\to \pi^+\eta') & = \lambda_{sd}\left(-\frac{1}{\sqrt{3}}\hat{T}_{18} \right) -\frac{\lambda_b}{2}\left(\frac{2}{\sqrt{3}}\bar{C}_{18} + \frac{1}{\sqrt{3}}\hat{T}_{18} \right)\,, \label{eq:su3-limit-amp-3}\\
        \mathcal{A}^{\text{SCS}}_{\text{SU(3)$_F$-lim.}}(D_s^+\to K^+\eta') & = \lambda_{sd}\left(\frac{1}{\sqrt{3}}\hat{T}_{18} \right) -\frac{\lambda_b}{2}\left(\frac{2}{\sqrt{3}}\bar{C}_{18} + \frac{1}{\sqrt{3}}\hat{T}_{18} \right)\,. \label{eq:su3-limit-amp-4}
\end{align}
Previously, when considering only the leading amplitude in Ref.~\cite{Bolognani:2024zno}, we noticed a decoupling of the neutral and charged modes as can be seen above. We note that in the $SU(3)_F$ limit, the subleading terms depend on a new $SU(3)_F$ limit parameter $\bar{C}_{18}$ that connects the charged and neutral modes. We discuss this parameter in more detail below.

From the amplitudes above,  the $SU(3)_F$-limit sum rules for the direct CP asymmetries follow
\begin{align}
        \left.a_{CP}^{\text{dir}}(D^0\to \pi^0\eta') \right|_{\text{SU(3)$_F$-lim.}}  & = -3\,\left.a_{CP}^{\text{dir}}(D^0\to \eta\eta') \right|_{\text{SU(3)$_F$-lim.}} \label{eq:su3limSR-1}\,,\\
        \left.a_{CP}^{\text{dir}}(D^+\to \pi^+\eta') \right|_{\text{SU(3)$_F$-lim.}}  & = -\,\left.a_{CP}^{\text{dir}}(D_s^+\to K^+\eta') \right|_{\text{SU(3)$_F$-lim.}} \,.\label{eq:su3limSR-2}
\end{align}
We can obtain a modified version of the sum rules, accounting for part of SU(3)$_F$-breaking effects in analogy to the sum rule given in Ref.~\cite{Bause:2022jes}. Concretely, starting from Eq.~(\ref{eq:aCPdir}), we write
\begin{align}
\frac{ a_{CP}^{\rm dir}(D^+ \rightarrow \pi^+ \eta') }{  a_{ CP}^{\rm dir}(D_s^+\to K^+ \eta')} &= \frac{ 
\mathrm{Im}\left(\mathcal{A}_b(D^+ \rightarrow \pi^+ \eta' )/\mathcal{A}_{sd}(D^+ \rightarrow \pi^+ \eta' )\right)
}{
\mathrm{Im}\left(\mathcal{A}_b(D_s^+\to K^+ \eta' )/\mathcal{A}_{sd}(D_s^+\to K^+ \eta' ) \right) 
}  \ ,
\end{align}
Using then the SU(3)$_F$-limit expression for $\mathcal{A}_b$ included in Table~\ref{tab:redefparams}, we find
\begin{align}
\frac{ a_{CP}^{\rm dir}(D^+ \rightarrow \pi^+ \eta') }{  a_{ CP}^{\rm dir}(D_s^+\to K^+ \eta')} 
&= \frac{ 
\mathrm{Im}\left( \overline{C}_{18}   /\mathcal{A}_{sd}(D^+ \rightarrow \pi^+ \eta' )\right)
}{
\mathrm{Im}\left(\overline{C}_{18} /\mathcal{A}_{sd}(D_s^+\to K^+ \eta' ) \right)
}  \\
&= \frac{
\vert \mathcal{A}_{sd}(D_s^+\to K^+ \eta' ) \vert
}{
\vert \mathcal{A}_{sd}(D^+ \rightarrow \pi^+ \eta' )\vert 
}
\frac{ 
\mathrm{Im}\left(\overline{C}_{18}  e^{-i \arg(\mathcal{A}_{sd}(D^+ \rightarrow \pi^+ \eta' ))}\right)
}{
\mathrm{Im}\left(\overline{C}_{18}  e^{-i \arg(\mathcal{A}_{sd}(D_s^+\to K^+ \eta' ) }\right)
} \,.
\end{align}

We remark that while the phase $\arg(\mathcal{A}_{sd})$ is unphysical, the relative phase between $\overline{C}_{18}$ and $\mathcal{A}_{sd}$ is physical.
Assuming the SU(3)$_F$ limit for this phase difference, gives (see Table~1 in Ref.~\cite{Bolognani:2024zno}) 
\begin{align}
\frac{ 
\mathrm{Im}\left(\overline{C}_{18}  e^{-i \arg(\mathcal{A}_{sd}(D^+ \rightarrow \pi^+ \eta' ))}\right)
}{
\mathrm{Im}\left(\overline{C}_{18}  e^{-i \arg(\mathcal{A}_{sd}(D_s^+\to K^+ \eta' ) }\right)
} &= -1\,.
\end{align}
Finally, we can express the $|\mathcal{A}_{sd}|$ amplitudes by the branching ratios by correcing for the phase space using Eq.~(\ref{eq:normalization}), we then find  
\begin{align}
\frac{  a_{CP}^{\rm dir}(D^+ \rightarrow \pi^+ \eta') }{  a_{ CP}^{\rm dir}(D_s^+\to K^+ \eta')} &=
 	-\sqrt{\frac{\mathcal{P}(D^+\rightarrow \pi^+ \eta')}{\mathcal{P}(D_s^+\rightarrow K^+ \eta')} }    
	\sqrt{\frac{\mathcal{B}(D_s^+ \to K^+ \eta')}{\mathcal{B}(D^+ \to \pi^+ \eta')} }    
	\,. \label{eq:improvedSR2} 
\end{align}
An equivalent derivation gives for the neutral modes:
\begin{align}
  \frac{ a_{ CP}^{\rm dir}(D^0\to \pi^0 \eta')}{ a_{CP}^{\rm dir}(D^0\to \eta \eta')} &= 
	-3 \sqrt{\frac{\mathcal{P}(D^0\rightarrow \pi^0 \eta')}{\mathcal{P}(D^0\rightarrow \eta\eta')}}     
	\sqrt{\frac{\mathcal{B}(D^0 \to \eta \eta')}{\mathcal{B}(D^0 \to \pi^0 \eta')}}     
	\, \label{eq:improvedSR1} \ .
\end{align}
Specifically, the sum rules Eqs.~(\ref{eq:improvedSR1}) and (\ref{eq:improvedSR2}) thus account for the 
SU(3)$_F$ breaking of the magnitude of the CKM-leading amplitude $\vert \mathcal{A}_{sd}\vert$ in Eq.~(\ref{eq:aCPdir}) and only apply the SU(3)$_F$-limit to $\mathcal{A}_b$ and to the relative strong phase between $\mathcal{A}_b$ and $\mathcal{A}_{sd}$. Eqs.~(\ref{eq:su3limSR-1}), (\ref{eq:su3limSR-2}) are readily obtained from Eqs.~(\ref{eq:improvedSR1}), (\ref{eq:improvedSR2}), respectively, when taking the SU(3)$_F$ limit for $\mathcal{A}_{sd}$.

\section{Predictions for CP Asymmetries \label{sec:predictions} }

\subsection{Numerical Inputs and Setup of the Analysis \label{sec:setup}} 

The general  setup of our frequentist statistical analysis 
is described in detail in the context of the branching ratio fit performed in Ref.~\cite{Bolognani:2024zno}.
In particular, we use the same branching ratio inputs as in Ref.~\cite{ParticleDataGroup:2024cfk}. 
Note that for consistency with our previous work we therefore do not include the recent additional measurement of $\mathcal{B}(D^+\to K^+\eta')$ from \cite{BESIII:2025ykz}, which is fully consistent with the world average.
In addition to the eight branching ratio data measurements, measurements of direct CP asymmetries are available for the two charged decay modes. 
The HFLAV~\cite{HeavyFlavorAveragingGroupHFLAV:2024ctg} averages are
\begin{align}
  \left.  a^{\text{dir}}_{CP}(D^+\to\pi^+\eta')\right|_{\text{exp}} &= (0.4\pm 0.2)\%\,, \quad 
	\text{\cite{HeavyFlavorAveragingGroupHFLAV:2024ctg, CLEO:2009fiz, Belle:2011tmj, LHCb:2022pxf}}
    \label{eq:piplus-cpasym} \\
    \left.a^{\text{dir}}_{CP}(D_s^+\to K^+\eta')\right|_{\text{exp}} &= (6.0\pm 18.9)\%\,. \quad
	\text{\cite{HeavyFlavorAveragingGroupHFLAV:2024ctg, CLEO:2009fiz}}
    \label{eq:Kplus-cpasym}
\end{align}
The measurement in Eq.~(\ref{eq:Kplus-cpasym}) has no constraining power due to 
its large uncertainty. Its very large central value would artificially influence the global minimum of our fit while only 
marginally constraining the involved parameters. Therefore, we do not include the measurement Eq.~(\ref{eq:Kplus-cpasym}) in our fit.  Note that we focus here on direct CP violation, while in the experiment, time-integrated or time-dependent measurements are performed.
In case of the neutral $D^0$ decays, such effects have to be subtracted from the experimental results~\cite{Gersabeck:2011xj, Kagan:2020vri,Nierste:2015zra}.

As in the branching ratio analysis of Ref.~\cite{Bolognani:2024zno}, we impose constraints on the size of $SU(3)_F$-breaking in the subleading amplitudes of all SCS decays
\begin{align}
\left|\frac{\mathcal{A}_{b,\cancel{\mathrm{{SU(3)}}}}(D\to P\eta')}{\mathcal{A}_{b,\mathrm{{SU(3)-lim}}}(D\to P\eta')}\right|\leq \varepsilon\,, 
\label{eq:SU3-breaking}
\end{align}
where $\varepsilon$ is the amount of $SU(3)_F$-breaking allowed. 
Note that consequently, we impose this constraint separately on both the CKM-leading and CKM-subleading part of the amplitudes.

We note that with future data it will be possible to test the assumption Eq.~(\ref{eq:SU3-breaking}) by probing the size of SU(3)$_F$ breaking effects with more precision. The measurements of CP asymmetries will also give information on possible cancellations in the SU(3)$_F$-limit $\mathcal{A}_b$ amplitude, see Eqs.~(\ref{eq:su3-limit-amp-1})--(\ref{eq:su3-limit-amp-4}). For example, if the contributions of $\overline{C}_{18}$ and $\hat{C}_{81}$ interfere destructively in the CP asymmetry of $D^0\rightarrow \pi^0\eta'$, they can not at the same time cancel in $D^0\rightarrow \eta\eta'$, see also Table~1. Such cancellations would appear in the data as suppressed CP asymmetries. Physics beyond the SM on the other hand will violate the resulting sum rules Eqs.~(\ref{eq:su3limSR-1}) and (\ref{eq:su3limSR-2}) at $\mathcal{O}(1)$.

The $\chi^2$ function is determined as
\begin{equation}
    \chi^2=(\vec{y}_{\text{data}}-\vec{y}_{\text{theo}})^T{\text{Cov}^{-1}}(\vec{y}_{\text{data}}-\vec{y}_{\text{theo}})\,,
\end{equation}
where $\vec{y}_{\text{data}}$ refers to the experimental input, ${\text{Cov}}$ is the covariance matrix for the experimental results and $\vec{y}_{\text{theo}}$ is the theoretical parametrisation of the respective observables based on Table~\ref{tab:redefparams}. We minimise this function using the Sequential Least SQuares Programming (SLSQP) algorithm implemented in SciPy \cite{kraft1988software,2020SciPy-NMeth}. For all results, minimisations are performed 100 times from random starting points for all parameters.

The $\Delta \mathcal{A}_b$ parameters of Table~\ref{tab:redefparams} are normalised to $|\hat{T}_{18}|$ and all phases are defined relative to the tree amplitude. 
The fit contains $17+8=25$ parameters 
for $8+1=9$ observables, 
with a total of $8+4=12$ $SU(3)_F$-breaking constraints on the amplitudes. 

A  $\chi^2$ scan of the allowed $SU(3)_F$ breaking $\varepsilon$ reproduces the pattern of Figure~2 of Ref.~\cite{Bolognani:2024zno}. In the limit of $SU(3)_F$ symmetry for both $\mathcal{A}_{sd}$ and $\Delta\mathcal{A}_b$, \emph{i.e.}~$\varepsilon=0$, the scan yields a $5.6\sigma$ discrepancy with respect to a perfect description of the data, $\chi^2=0$, which is reached for $\varepsilon=50\%$, consistent with the branching ratio only fit and confirming the exclusion of the $SU(3)_F$ limit. At $30\%$ $SU(3)_F$-breaking, the same $\chi^2=6.01$ of the branching ratio fit is obtained, which
indicates that the additional CP violating parameters absorb the effect of the single experimental result. 
We note that the non-vanishing value of $\chi^2$ at the global minimum complies with the expected statistical fluctuations of the branching ratio measurements, see Fig.~4 in Ref.~\cite{Bolognani:2024zno}.
In the following, we fix 
\begin{align}
\varepsilon=30\%\,.
\end{align}
in all fits and calculate the $\Delta \chi^2$ with respect to the corresponding global minimum.

We stress that the physical constraints on SU(3)$_F$-breaking that we impose on the fit always apply to parts of amplitudes, not to single parameters. 
In that way, our fit results remain independent of the chosen reparametrisation of parameters.

For technical reasons, we include very wide bounds on the SU(3)$_F$-breaking parameters $\mathcal{T}^{(1)}_i$.  These bounds have no physical meaning and are just a technical choice when performing the fit.
We checked that our results are not touching this boundary, and are therefore independent of this choice.

\begin{figure}[t]
    \centering
    \includegraphics[width=0.6\textwidth]{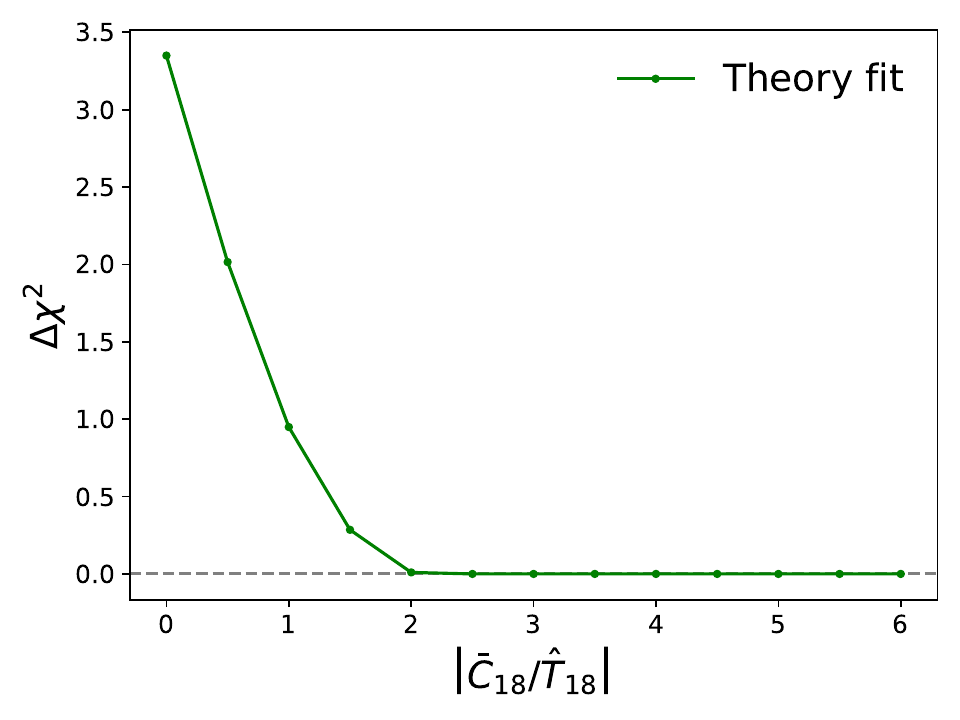}
    \caption{%
        $\Delta \chi^2$ profile of $|\bar{C}_{18}/\hat{T}_{18}|$.}
    \label{fig:plateau-c18}
\end{figure}

The SU(3)$_F$ limit parameter $\bar{C}_{18}$, which is a combination of color-suppressed tree and penguin diagrams, plays an important role for the prediction of the considered CP asymmetries, as it is the only SU(3)$_F$ limit parameter involved which is not constrained by branching ratio measurements.
In Fig.~\ref{fig:plateau-c18} we show its $\Delta \chi^2$ profile. Starting from $|\bar{C}_{18}/\hat{T}_{18}|\sim 2$, we observe a plateau with $\Delta\chi^2=0$, \emph{i.e.}, a perfect accommodation of the CP violation data point.
This is a reflection of the lack of data on CP asymmetries to constrain the fit: For a significantly large enough $\bar{C}_{18}$, its relative phase to the CKM-leading amplitude can always be tuned such that the single measured CP asymmetry is perfectly described. 
At smaller magnitudes, however, the $\Delta\chi^2$ increases, with $|\bar{C}_{18}/\hat{T}_{18}|=1$ yielding $\Delta \chi^2 = 0.95$.
The expectation from a $1/N_c$ power counting~\cite{Buras:1985xv, tHooft:1973alw} would be $|\bar{C}_{18}/\hat{T}_{18}|=1/3$, which gives $\Delta \chi^2 = 2.45$, corresponding to a tension with the data at $1.6 \sigma$.

\subsection{Results}

\begin{table}[t]
    \centering
    \begin{tabular}{ccc}\hline\hline
        Observable  & Theory fit     & Experiment~\cite{HeavyFlavorAveragingGroupHFLAV:2024ctg}  \\\hline
        $a_{CP}^{\mathrm{dir}}(D^0\to \pi^0\eta' )$  & $[-0.66,0.66]\%$ & \textemdash  \\
         $a_{CP}^{\mathrm{dir}}(D^0\to \eta \eta')$  & $[-0.29,0.29]\%$ & \textemdash  \\
         $a_{CP}^{\mathrm{dir}}(D^+\to \pi^+\eta' )$ & $(0.40_{-0.20}^{+0.19})\%$ & $(0.40 \pm 0.20)\%$ \\
         $a_{CP}^{\mathrm{dir}}(D_s^+\to K^+\eta')$  & $[-0.62,0.36]\%$ &  $(6.0\pm 18.9)\%^a$   \\\hline\hline
    \end{tabular}
    \caption{Predictions for direct CP asymmetries from our global theory fit, compared to the experimental input data. In cases with large non-Gaussianities we show the $1\sigma$ ranges and no central value. $^a$Not included in the fit, see Sec.~\ref{sec:setup} for details.}  
    \label{tab:resultsCP}
\end{table}

\begin{figure}[t]
\subfigure[]{\includegraphics[width=0.49\textwidth]{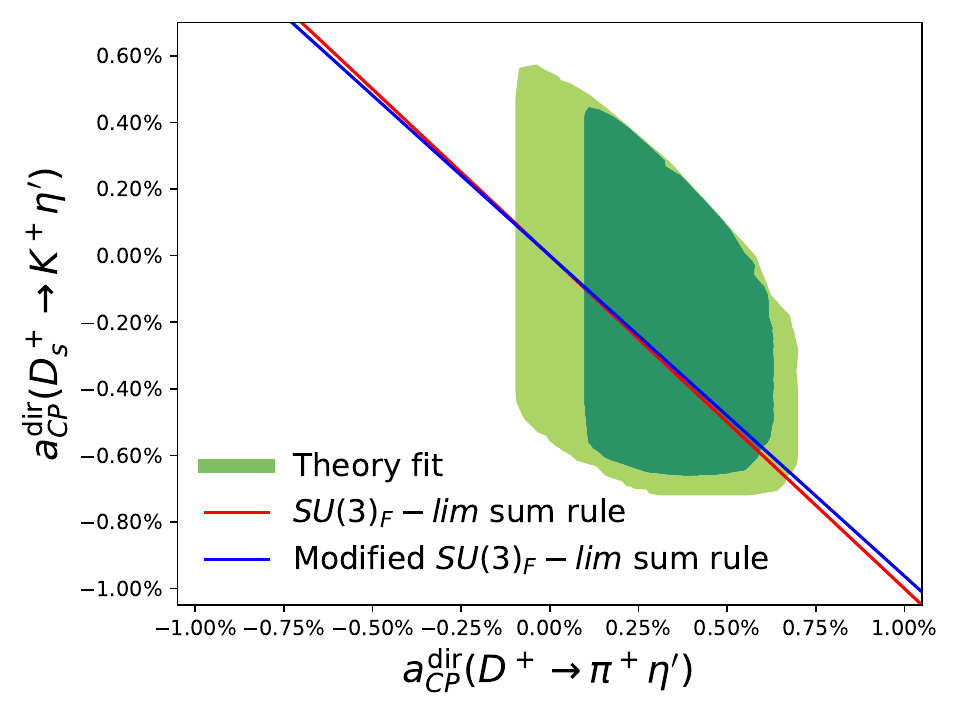}}
\subfigure[]{\includegraphics[width=0.49\textwidth]{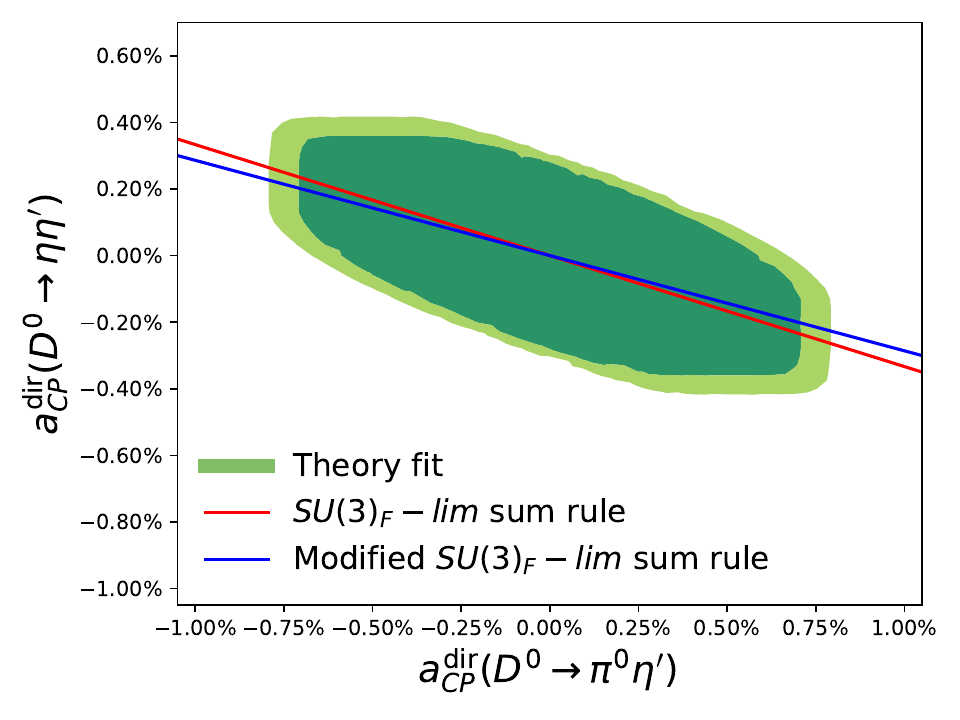}}
    \caption{
        Correlations between predicted CP asymmetries for the charged (left) and neutral (right) modes. 
	The red line represents the $SU(3)_F$ limit sum rules Eqs.~(\ref{eq:su3limSR-1}, \ref{eq:su3limSR-2}). 
	The blue line indicates the modified sum rules Eqs.~(\ref{eq:improvedSR1}, \ref{eq:improvedSR2}) which partially account for SU(3)$_F$-breaking effects. The contour lines represent the $68.30\%$ and $95.45\%$ confidence levels. 
    }
    \label{fig:correlations-Cpasyms}
\end{figure}

\begin{figure}[t]
    \centering
\subfigure[]{\includegraphics[width=0.49\textwidth]{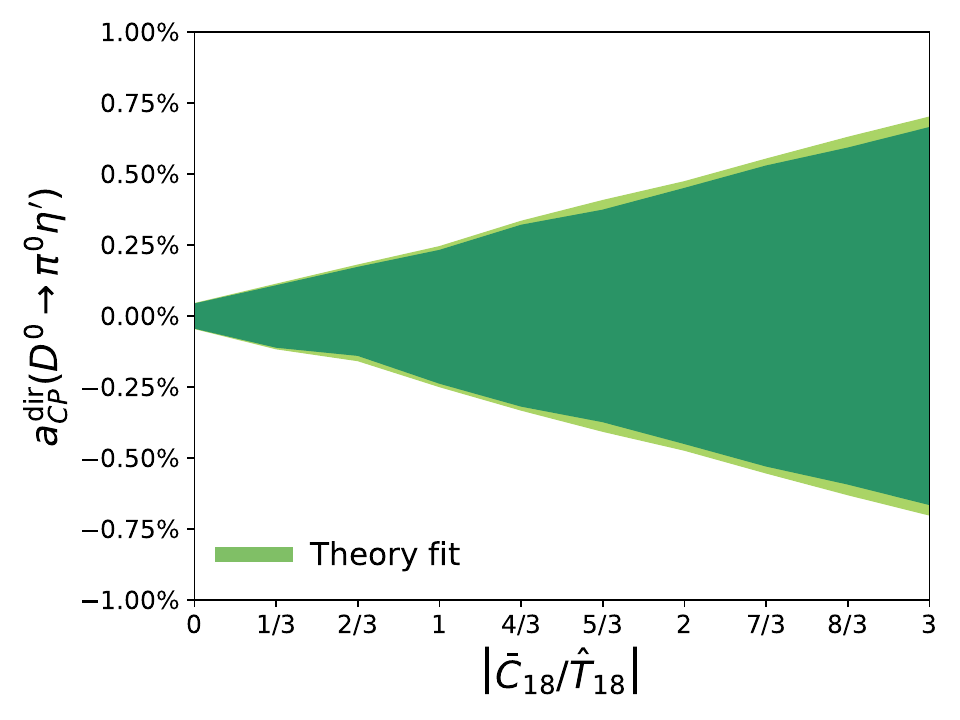}}
\subfigure[]{\includegraphics[width=0.49\textwidth]{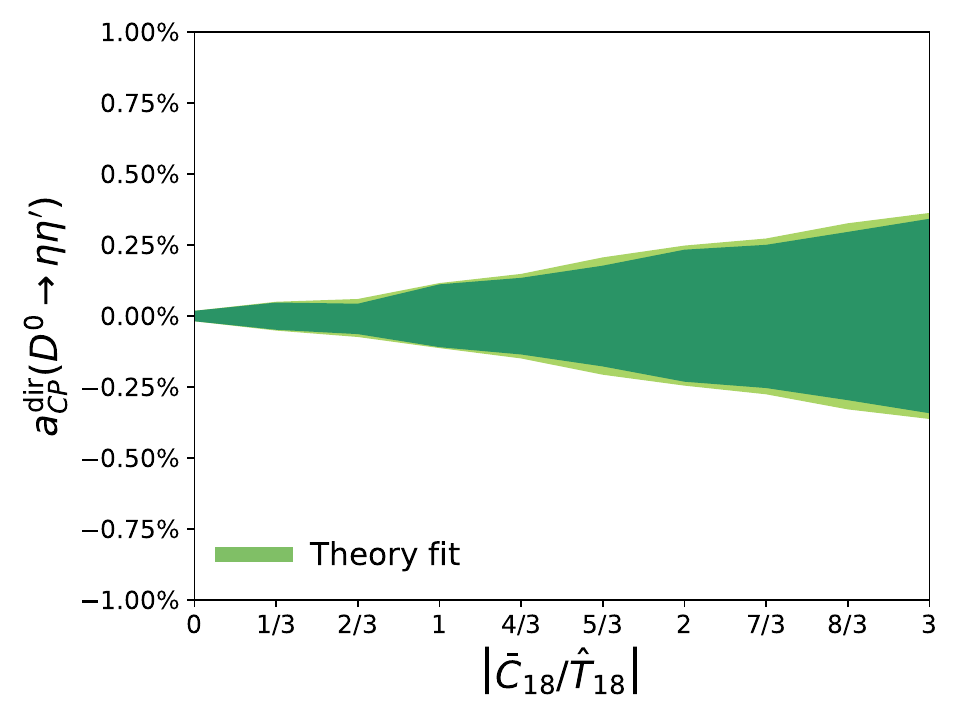}}\\
\subfigure[\label{fig:subc}]{\includegraphics[width=0.49\textwidth]{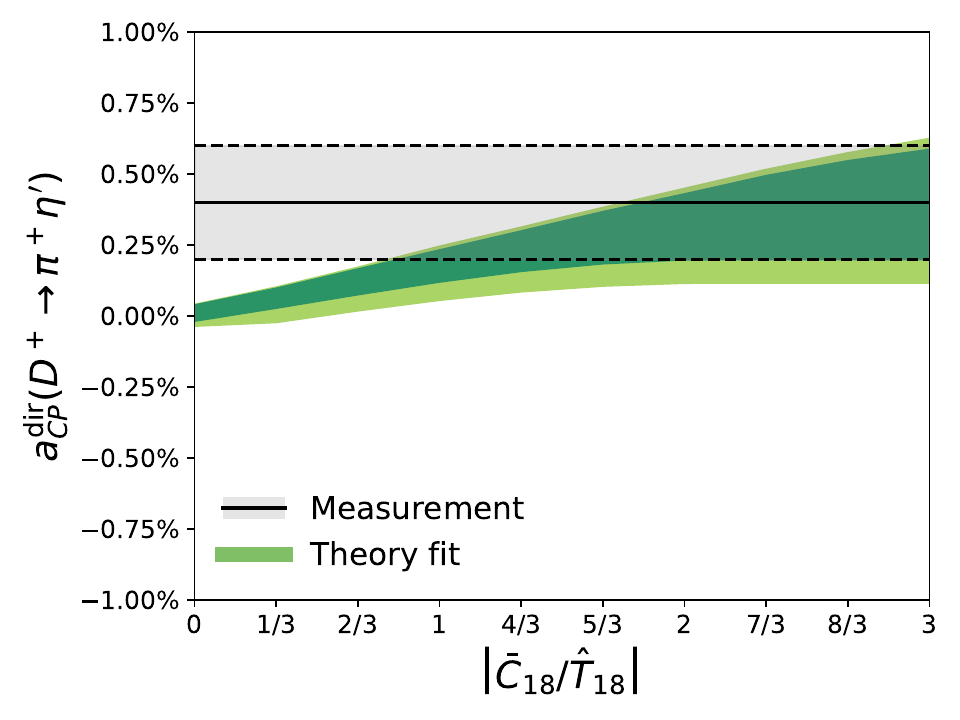}}
\subfigure[]{\includegraphics[width=0.49\textwidth]{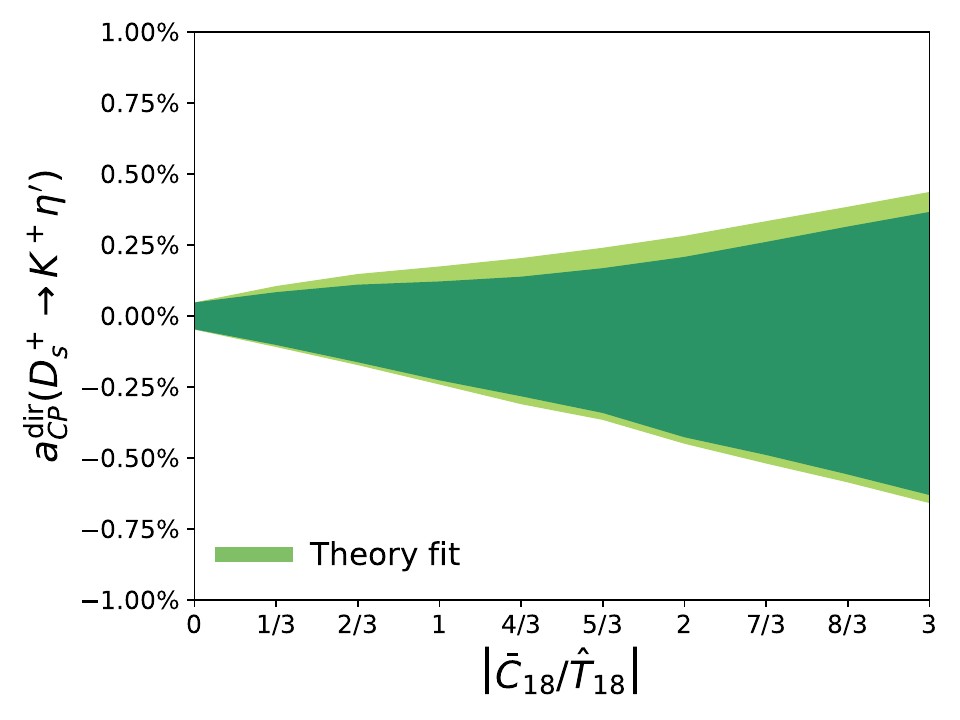}}
    \caption{%
        Predictions for CP asymmetries as a function of the SU(3)$_F$ limit parameter ratio $\left| \bar{C}_{18} / \hat{T}_{18} \right|$. The filled green areas represent the $1\sigma$ and $2\sigma$ confidence levels. The gray band in $a_{CP}^{\textrm{dir}}(D^+\to\pi^+\eta')$ corresponds to the $1\sigma$ experimental input. 
	The plots can be used as a tool in order to interpret future determinations of CP asymmetries in terms of allowed intervals for $\left| \bar{C}_{18} / \hat{T}_{18} \right|$ and check their consistency.
    }
    \label{fig:cpfixedc18}
\end{figure}
In order to make predictions for the CP asymmetries, it is necessary to impose a constraint on $\bar{C}_{18}$ because the single measured CP asymmetry is not sufficient to constrain this parameter. Using
\begin{align}
   \left| \frac{\bar{C}_{18}}{\hat{T}_{18}}\right| \leq 3\,, \label{eq:1Nc-assumption}
\end{align}   
we obtain the results of our theory fit for the CP asymmetries given in Table~\ref{tab:resultsCP} along with the available experimental data. As discussed below, future experimental measurements will make it possible to systematically probe this assumption using our results.

The imposed theoretical constraints Eq.~(\ref{eq:SU3-breaking}) and Eq.~(\ref{eq:1Nc-assumption}) partially lead to a non-Gaussian $\Delta \chi^2$-profiles, 
featuring flat directions and steep ascends of the $\Delta\chi^2$. In such cases we quote intervals rather than a central value plus error. We note that we predict CP asymmetries at the sub-percent level and perfectly reproduce the experimental value for $D^+\to \pi^+\eta'$. 

Interestingly, the underlying approximate $SU(3)_F$ symmetry results in correlations between the CP asymmetries. 
In Fig.~\ref{fig:correlations-Cpasyms} we show the correlation obtained in our theory fit between charged and neutral modes. In addition, we show the correlation imposed by the 
sum rules Eqs.~(\ref{eq:su3limSR-1})--(\ref{eq:improvedSR2}) in the SU(3)$_F$ limit.
Our analysis indicates how far the CP asymmetries can deviate from these sum rules once SU(3)$_F$-breaking contributions are taken into account. If future measurements deviate from these correlations, or find individual CP asymmetries beyond our predictions, this could indicate that $\vert \bar{C}_{18} / \hat{T}_{18}\vert$ is larger than our bound in Eq.~(\ref{eq:1Nc-assumption}). This ratio probes the CKM-subleading effects over the CKM-leading effects (also referred to as the penguin over tree ratio). Such a violation of our bound would already be interesting in itself. 
Hints for a corresponding related enhancement have been seen in the  $D\rightarrow \pi\pi$ system, see~\emph{e.g.}~the fit results for the ratio of penguin over tree diagrams with central values of $\sim 5$ found in Ref.~\cite{Gavrilova:2023fzy}, triggered by the latest results from LHCb on $a_{CP}^{dir}(D^0\rightarrow \pi^+\pi^-)$ \cite{LHCb:2022lry}. We stress that in our case, due to the reparametrization Eq.~(\ref{eq:redef-penguin}), it is not possible to disentangle the color-suppressed tree from the penguin. This also means that for $D\to P \eta'$, we would obtain non-vanishing CP asymmetries even in the limit where all penguin diagrams are zero.

New measurements of the CP asymmetries are therefore of key importance. Such future measurements could then be used to extract the parameter $\vert \bar{C}_{18} / \hat{T}_{18}\vert$ directly from data.
In order to facilitate this extraction we provide in Fig.~\ref{fig:cpfixedc18} the allowed ranges for the CP asymmetries as a function of $\vert\bar{C}_{18} / \hat{T}_{18}\vert$. We stress that the assumption Eq.~(\ref{eq:1Nc-assumption}) does not enter Fig.~\ref{fig:cpfixedc18}. Once a CP asymmetry is measured, the corresponding interval of $\vert \bar{C}_{18} / \hat{T}_{18}\vert$ can be read off from the corresponding plot. 
Several CP asymmetry measurements would then allow to check for consistency between the different modes and the validity of our $SU(3)_F$-based analysis.  

In particular, Figure \ref{fig:subc} shows that the current data for $a_{CP}^{dir}(D^+\rightarrow \pi^+ \eta')$ prefers 
\begin{align}
\vert \bar{C}_{18}/\hat{T}_{18}\vert\gtrsim 1\,.
\end{align}
With future more precise data on $a_{CP}^{dir}(D^+\rightarrow \pi^+ \eta')$, it will be possible to probe the expectation from $1/N_c$ power counting $\vert \bar{C}_{18} / \hat{T}_{18}\vert \sim 0.3$.

\section{Conclusions \label{sec:conclusions}}

The direct CP asymmetries of SCS $D\rightarrow P\eta'$ decays are basically uncharted territory and offer a big opportunity to further explore CP violation in the charm sector. We provide SM predictions based on the approximate SU(3)$_F$ symmetry of QCD including first order breaking effects that allow the interpretation of future data. In our global analysis we allow for $30\%$ linear $SU(3)_F$-breaking effects. 

The major results of this paper are the correlations between the four CP asymmetries obtained from our global theory fit to the available $D\rightarrow P\eta'$ data in  Fig.~\ref{fig:correlations-Cpasyms}. These plots show how far the CP asymmetries can deviate from their SU(3)$_F$ limit sum rules, and serve as a future test of the consistency of the data. 

Compared to the CKM-leading amplitude relevant for branching ratio studies, the CKM-subleading part, crucial for CP violation, is described by one additional $SU(3)_F$ limit parameter $\bar{C}_{18}$ which also contains the penguin topologies. 
The current measurement of the $D^+\to \pi^+\eta'$ CP asymmetry places a weak constraint on the ratio $\vert \bar{C}_{18} / \hat{T}_{18}\vert \gtrsim 1$, which probes the CKM-subleading over CKM-leading amplitudes. Further constraining this ratio from experimental data is key to understanding CP violation in charm, not only in $D\to P\eta'$ but also beyond, and to probe expectations based on the $1/N_c$ power counting. Our predictions displayed in Fig.~\ref{fig:cpfixedc18} can be used to extract this ratio directly from data, allowing for the early identification of possible deviations from the Standard Model.

\acknowledgments

The work of K.K.V. is supported in part by the Dutch Research Council (NWO) as part of the project Solving
Beautiful Puzzles (VI.Vidi.223.083) of the research programme Vidi. 
S.S. is supported by the STFC through an Ernest Rutherford Fellowship under reference ST/Z510233/1 and the grant ST/X003167/1.

\appendix

\section{Topological Decomposition without Redefinitions \label{sec:tables}}

\begin{figure}[t]
    \centering

\subfigure[ $P_{88}^{(1)}$]{\includegraphics[width=0.24\textwidth]{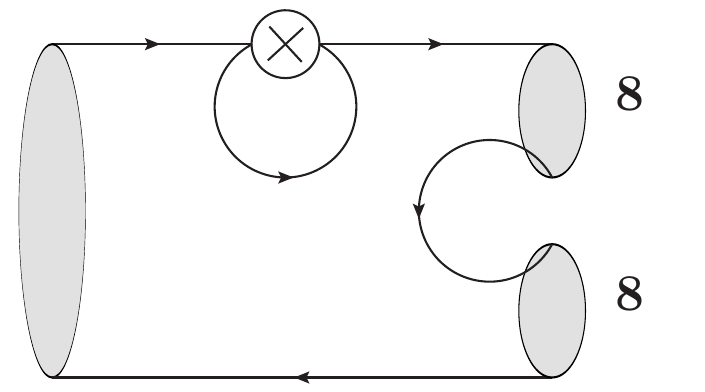}}
\subfigure[ $P_{11}^{(1)}$]{\includegraphics[width=0.24\textwidth]{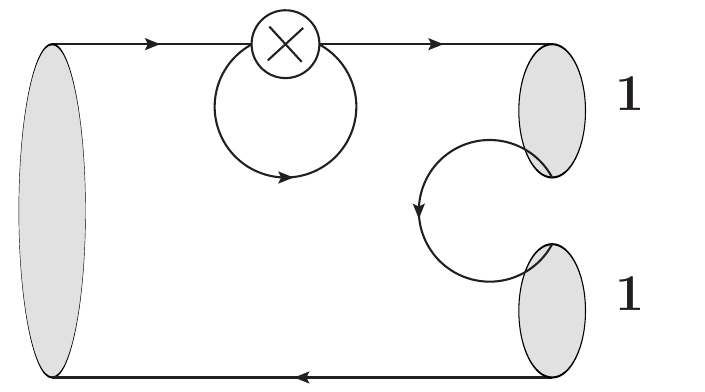}}
\subfigure[ $P_{ij,1}^{(1)}$]{\includegraphics[width=0.21\textwidth]{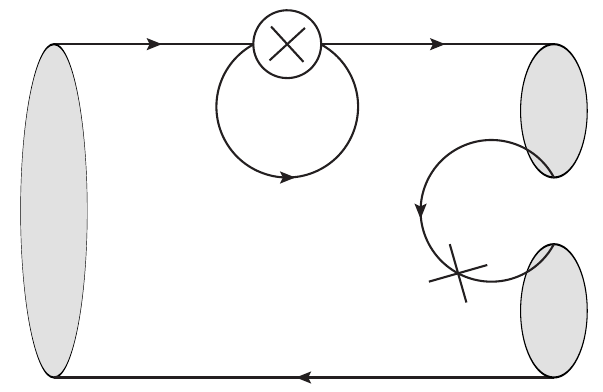}}
\subfigure[ $P_{ij,2}^{(1)}$]{\includegraphics[width=0.21\textwidth]{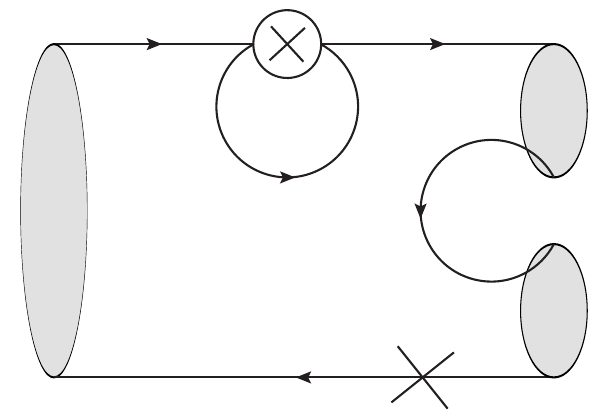}}
\subfigure[ $P_{H,11}^{(1)}$]{\includegraphics[width=0.24\textwidth]{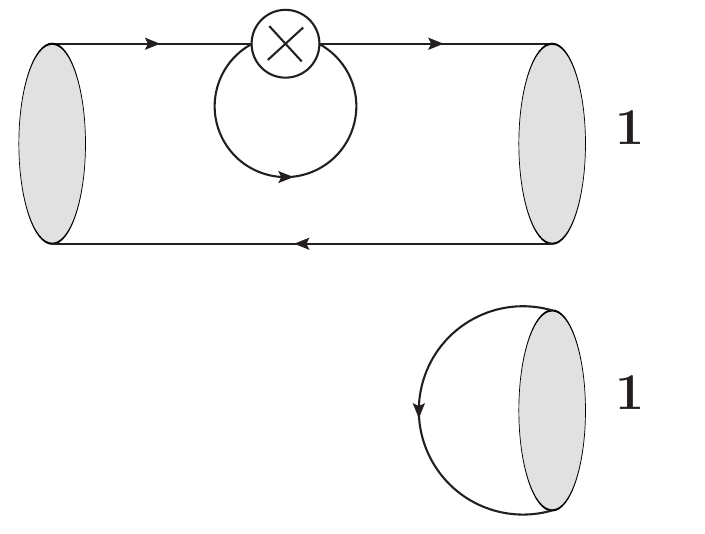}}
\subfigure[ $P_{H,1}^{(1)}$]{\includegraphics[width=0.21\textwidth]{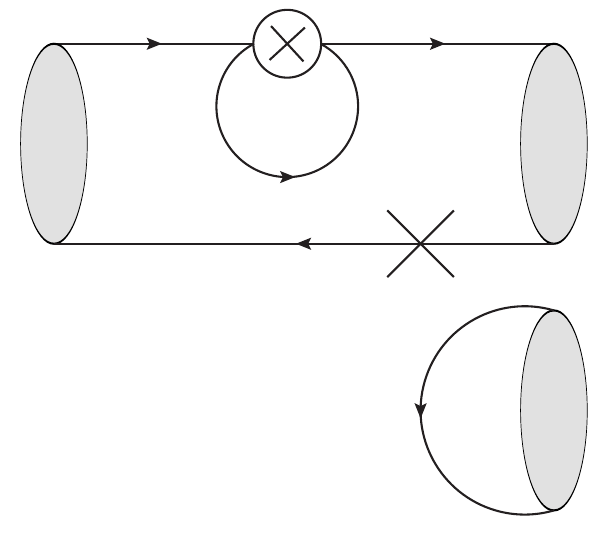}}\\
\subfigure[ $PA_{88}^{(1)}$]{\includegraphics[width=0.24\textwidth]{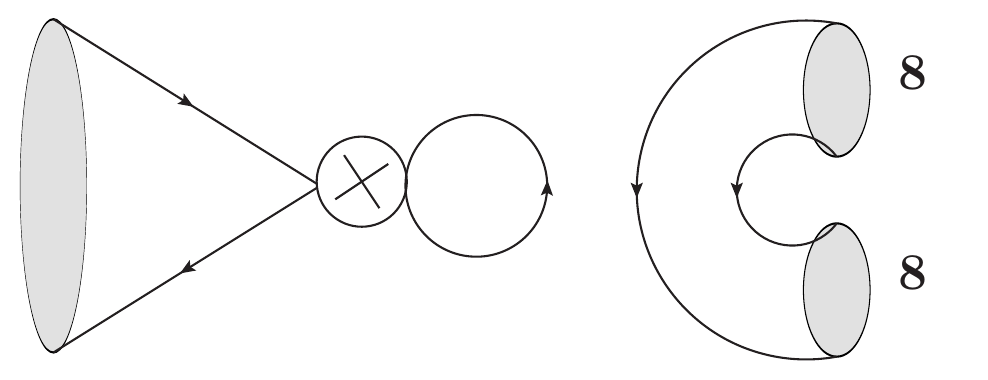}}
\subfigure[ $PA_{11}^{(1)}$]{\includegraphics[width=0.24\textwidth]{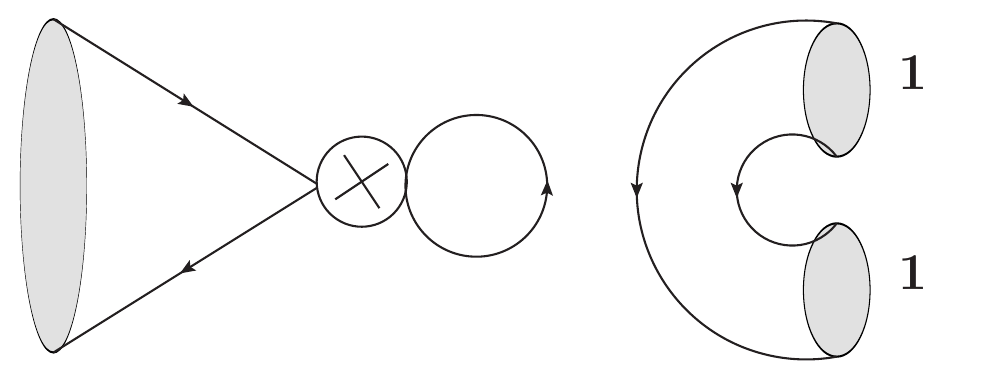}}
\subfigure[ $PA_{ij,1}^{(1)}$]{\includegraphics[width=0.215\textwidth]{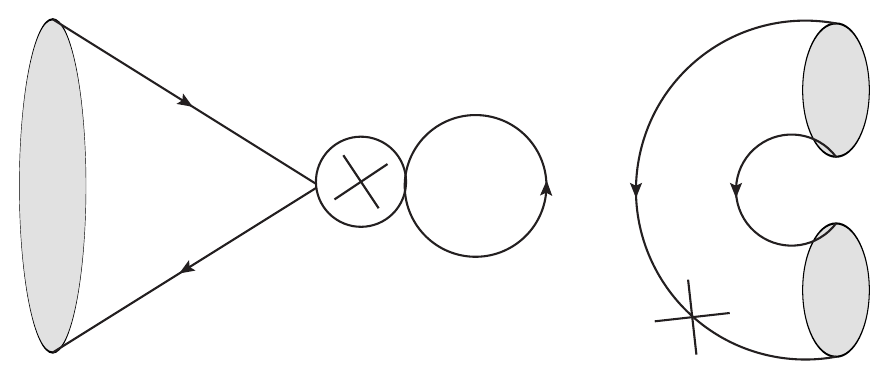}}
\subfigure[ $PA_{ij,2}^{(1)}$]{\includegraphics[width=0.215\textwidth]{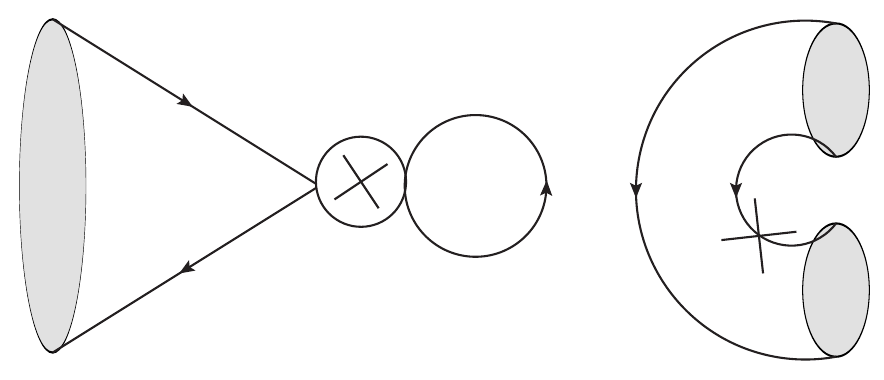}}
\subfigure[ $PA_{H,11}^{(1)}$]{\includegraphics[width=0.24\textwidth]{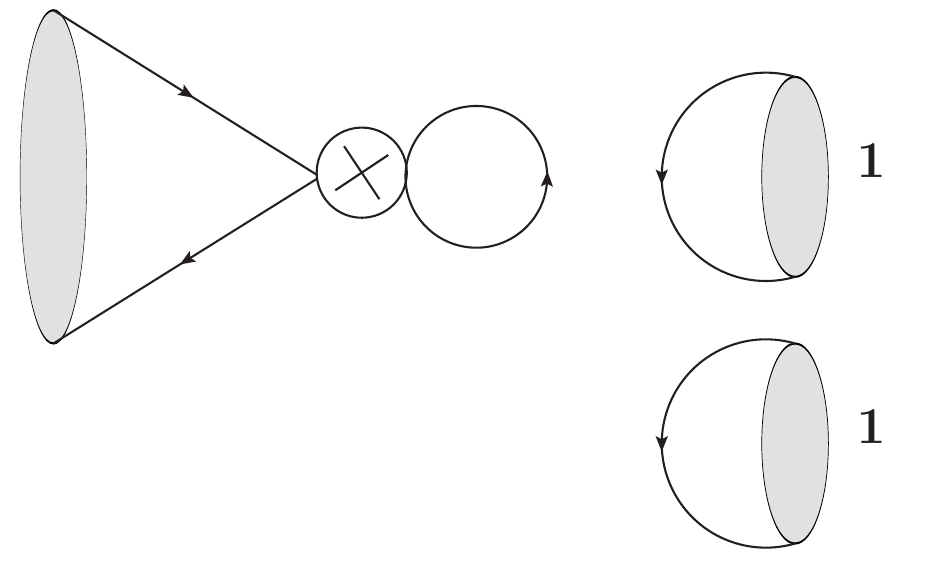}}
\subfigure[ $PA_{H,1}^{(1)}$]{\includegraphics[width=0.21\textwidth]{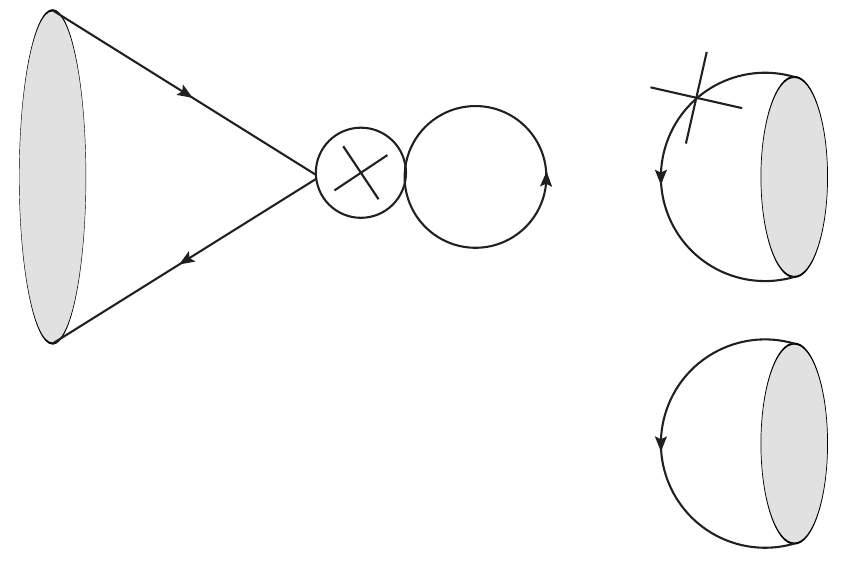}}\\
\subfigure[ $C_{11}^{(1)}$]{\includegraphics[width=0.24\textwidth]{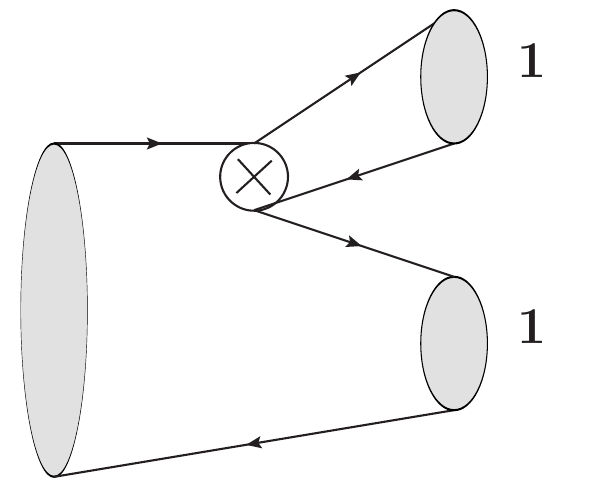}}
\subfigure[ $E_{11}^{(1)}$]{\includegraphics[width=0.24\textwidth]{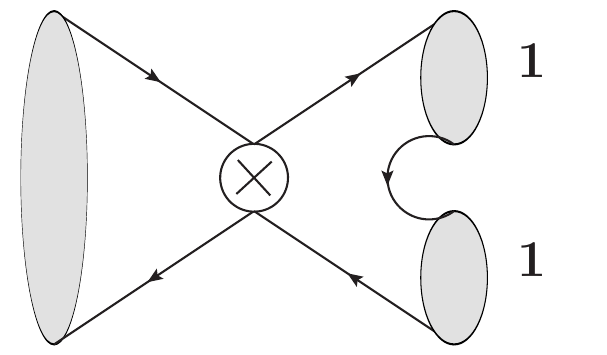}}
\subfigure[ $E_{H,11}^{(1)}$]{\includegraphics[width=0.24\textwidth]{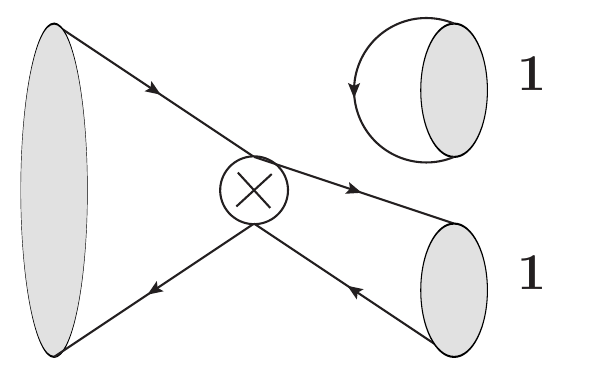}}
    \caption{%
        $SU(3)_F$-breaking topologies contributing to $\mathcal{A}_{b}$. The subscripts $ij$ indicate the mixed singlet-octet contributions, \emph{i.e.}, $18$ or $81$. The Feynman diagrams are created with JaxoDraw~\cite{Binosi:2003yf, Binosi:2008ig}.
    }
    \label{fig:Absuppressedtopologies}
\end{figure}

In Tables~\ref{tab:Ab-Ts-As}, \ref{tab:Ab-Cs-Es} and \ref{tab:Ab-Ps} we show the topological decomposition of the CKM-subleading amplitudes $\mathcal{A}_b(d)$, complementing the 
CKM-leading ones given in Ref.~\cite{Bolognani:2024zno}. We show the definitions of the topologies in Fig.~\ref{fig:Absuppressedtopologies}.

\newpage
\begin{table}[t]
\begin{center}
\setlength{\tabcolsep}{0.5pt} 
\begin{tabular} 
{c|cccc|cccccccccccc}
\hline \hline
Decay ampl. $\mathcal{A}_b(d)$ & 
$T_{18}$ &   
$A_{18}$ &  $A_{81}$ & $A_{H}$ &
 $T_{18,1}^{(1)}$ & $T_{18,2}^{(1)}$ & $T_{18,3}^{(1)}$ &  
 $T_{88}^{(1)}$ &
 $A_{18,1}^{(1)}$ & $A_{18,2}^{(1)}$ & 
 $A_{81,1}^{(1)}$ & $A_{81,2}^{(1)}$ & $A_{81,3}^{(1)}$ & 
 $A_{88}^{(1)}$ &
 $A_{H,1}^{(1)}$& $A_{H,2}^{(1)}$ 
 \\\hline\hline
$D^0 \rightarrow \pi^0 \eta'$  & 0 & 0 & 0 & 0 & 
$0$ & $0$ & $0$ &  
$0$ &
$0$ & $0$ &  
$0$ & $0$ & $0$ & 
$0$ &
  $0$ & $0$ \\\hline\hline
$D^0 \rightarrow \eta \eta'$  & 0 & 0 & 0 & 0 & 
$0$ & $0$ & $0$ &  
$0$ &
$0$ & $0$ &  
$0$ & $0$ & $0$ & 
$0$ &
  $0$ & $0$ \\\hline\hline
$D^+ \rightarrow \pi^+ \eta'$ &  $\frac{1}{\sqrt{3}}$ & $\frac{1}{\sqrt{3}}$ & $\frac{1}{\sqrt{3}}$ & $\frac{3}{\sqrt{3}}$ & 
$0$ & $0$ & $0$ &  
$\frac{1}{\sqrt{6}}$ &
$0$ & $0$ &  
$0$ & $0$ & $0$ & 
$\sqrt{\frac{2}{3}}$ &
  $0$ & $0$
\\\hline\hline
$D_s^+ \rightarrow K^+ \eta'$ &  $\frac{1}{\sqrt{3}}$ & $\frac{1}{\sqrt{3}}$ & $\frac{1}{\sqrt{3}}$ & $\frac{3}{\sqrt{3}}$ & 
$\frac{1}{\sqrt{3}}$ & $\frac{1}{\sqrt{3}}$ & $\frac{1}{\sqrt{3}}$ & 
$-\sqrt{\frac{2}{3}}$ &
$\frac{1}{\sqrt{3}}$ &  $\frac{1}{\sqrt{3}}$ &  
$\frac{1}{\sqrt{3}}$ &  $\frac{1}{\sqrt{3}}$ & $\frac{1}{\sqrt{3}}$ & 
$-\frac{1}{\sqrt{6}}$ &
 $\sqrt{3}$ &  $\sqrt{3}$  \\\hline\hline
\end{tabular}
\caption{Decomposition of $\mathcal{A}_b$ coefficients for tree ($T$) and annihilation ($A$) topologies.} 
\label{tab:Ab-Ts-As}
\end{center}
\end{table}

\begin{table}[t]
\begin{tiny}
\begin{center}
\setlength{\tabcolsep}{0.5pt} 
\resizebox{1.1\linewidth}{!}{%
\hspace*{-1cm}\begin{tabular} 
{c|ccccc|cccccccccccccccccc}
\hline \hline
Decay ampl. $\mathcal{A}_b(d)$ & 
$C_{18}$ & $C_{81}$ & 
$E_{18}$ & $E_{81}$ & $E_{H}$ & $C_{18,1}^{(1)}$ & $C_{18,2}^{(1)}$ &
 $C_{18,3}^{(1)}$ &
 $C_{81,1}^{(1)}$ & $C_{81,2}^{(1)}$ & 
 $C_{88}^{(1)}$ &
 $C_{11}^{(1)}$ &
 $E_{18,1}^{(1)}$ & $E_{18,2}^{(1)}$ & $E_{18,3}^{(1)}$ & 
 $E_{81,1}^{(1)}$ & $E_{81,2}^{(1)}$ & $E_{81,3}^{(1)}$ & 
 $E_{88}^{(1)}$ &
 $E_{11}^{(1)}$ &
 $E_{H,1}^{(1)}$ & $E_{H,2}^{(1)}$ & $E_{H,11}^{(1)}$
 \\\hline\hline
$D^0 \rightarrow \pi^0 \eta'$  & $\frac{2}{\sqrt{6}}$ & $-\frac{1}{\sqrt{6}}$ & $-\frac{1}{\sqrt{6}}$ & $-\frac{1}{\sqrt{6}}$ & $-\frac{3}{\sqrt{6}}$ &
$\frac{1}{\sqrt{6}}$ & $\frac{1}{\sqrt{6}}$ &
 $0$ &
 $0$ & $0$ &
 $-\frac{1}{\sqrt{3}}$ & $0$ &
 $0$ & $0$ & $0$ & 
 $0$ & $0$ & $0$ & 
 $-\frac{1}{\sqrt{3}}$ &
 $0$ & $0$ & $0$ & $0$
\\\hline\hline
$D^0 \rightarrow \eta \eta'$  & $\frac{2}{\sqrt{18}}$ & $-\frac{1}{\sqrt{18}}$ & $-\frac{1}{\sqrt{18}}$ & $-\frac{1}{\sqrt{18}}$ & $-\frac{3}{\sqrt{18}}$ &
$\frac{1}{3\sqrt{2}}$ & $\frac{1}{3\sqrt{2}}$ & $0$ & 
 $-\frac{\sqrt{2}}{3}$ & $-\frac{\sqrt{2}}{3}$ & 
 $-\frac{1}{3}$ &  $\frac{4}{3}$ &
 $-\frac{\sqrt{2}}{3}$ & $-\frac{\sqrt{2}}{3}$ & $-\frac{\sqrt{2}}{3}$ & 
 $-\frac{\sqrt{2}}{3}$ & $-\frac{\sqrt{2}}{3}$ & $-\frac{\sqrt{2}}{3}$ & 
 $\frac{5}{3}$ & $\frac{4}{3}$ &
 $-\sqrt{2}$ & $-\sqrt{2}$ & $4$
\\\hline\hline
$D^+ \rightarrow \pi^+ \eta'$ &  $\frac{2}{\sqrt{3}}$ & 0 & 0 & 0 & 0 & $\frac{1}{\sqrt{3}}$ & $\frac{1}{\sqrt{3}}$ &  
 $0$ & $0$ & $0$ &
 $-\frac{1}{\sqrt{6}}$ &  $0$ &
 $0$ & $0$ &  $0$ &
 $0$ & $0$ & $0$ & 
 $0$ &
 $0$ & $0$ & $0$ & $0$ 
\\\hline\hline
$D_s^+ \rightarrow K^+ \eta'$ &  $\frac{2}{\sqrt{3}}$ & 0 & 0 & 0 & 0 & $\frac{1}{\sqrt{3}}$ & $\frac{1}{\sqrt{3}}$ & 
 $\frac{2}{\sqrt{3}}$ &
 $0$ & $0$ & 
 $-\frac{1}{\sqrt{6}}$ & $0$ &
 $0$ & $0$ & $0$ & 
 $0$ & $0$ & $0$ & 
 $0$ &
 $0$ & $0$ & $0$ & $0$  \\\hline\hline
\end{tabular}}
\caption{Decomposition of $\mathcal{A}_b$ coefficients for colour-suppressed trees ($C$) and exchange ($E$) topologies. } 
\label{tab:Ab-Cs-Es}
\end{center}
\end{tiny}
\end{table}

\begin{table}[t]
\begin{tiny}
\begin{center}
\setlength{\tabcolsep}{0.5pt} 
\resizebox{1.1\linewidth}{!}{%
\hspace*{-1cm}\begin{tabular} %
{c|ccc|ccccccccccccccc}
\hline \hline
Decay ampl. $\mathcal{A}_b(d)$ & 
$P_{18}$ & $P_{81}$ & $P_{H}$
& $P_{18,2}^{(1)}$ & $P_{81,1}^{(1)}$ & $P_{81,2}^{(1)}$ &  $P_{88}^{(1)}$ & $P_{11}^{(1)}$ &
 $P_{H,1}^{(1)}$ & $P_{H,11}^{(1)}$ &
 $PA_{18,1}^{(1)}$ & $PA_{18,2}^{(1)}$ & $PA_{81,1}^{(1)}$ & $PA_{81,2}^{(1)}$ & $PA_{88}^{(1)}$ &   $PA_{11}^{(1)}$ & $PA_{H,1}^{(1)}$ & $PA_{H,11}^{(1)}$
 \\\hline\hline
$D^0 \rightarrow \pi^0 \eta'$  & $\frac{1}{\sqrt{6}}$ & $\frac{1}{\sqrt{6}}$ & $\frac{3}{\sqrt{6}}$ & 0 & 0 & 0 & $\frac{1}{\sqrt{3}}$ & 0 & 0 & 0 & 0 & 0 & 0 & 0 & 0 & 0 & 0 & 0 
\\\hline\hline
$D^0 \rightarrow \eta \eta'$  & 
$\frac{1}{\sqrt{18}}$ & $\frac{1}{\sqrt{18}}$ & $\frac{1}{\sqrt{2}}$ & 0 & 0 & 0 & $\frac{1}{3}$ & $\frac{2}{3}$ & 0 & $2$ &  -$\frac{\sqrt{2}}{3}$ & -$\frac{\sqrt{2}}{3}$ & -$\frac{\sqrt{2}}{3}$ & -$\frac{\sqrt{2}}{3}$ & $2$ &  $2$ & $-\sqrt{2}$ & $6$
\\\hline\hline
$D^+ \rightarrow \pi^+ \eta'$ & 
$\frac{1}{\sqrt{3}}$ & $\frac{1}{\sqrt{3}}$ & $\frac{3}{\sqrt{3}}$ & 0 & 0 & 0 & $\frac{2}{\sqrt{6}}$ & 0 & 0 & 0 & 0 & 0 & 0 & 0 & 0 & 0 & 0 & 0
\\\hline\hline
$D_s^+ \rightarrow K^+ \eta'$ &  
$\frac{1}{\sqrt{3}}$ & $\frac{1}{\sqrt{3}}$ & $\frac{3}{\sqrt{3}}$ & $\frac{1}{\sqrt{3}}$ & $\frac{1}{\sqrt{3}}$ & $\frac{1}{\sqrt{3}}$ & $-\frac{1}{\sqrt{6}}$ & 0 & $\frac{3}{\sqrt{3}}$ & 0 & 0 & 0 & 0 & 0 & 0 & 0 & 0 & 0
\\\hline\hline
\end{tabular}}
\caption{Decomposition of $\mathcal{A}_b$ coefficients for penguin topologies ($P$ and $PA$, for penguin-annihilation). } 
\label{tab:Ab-Ps}
\end{center}
\end{tiny}
\end{table}

\clearpage

\bibliographystyle{JHEP}
\bibliography{draft.bib}

@article{Lindenbaum:1980rf,
    author = "Lindenbaum, S. J.",
    title = "{The OZI rule and glueballs}",
    reportNumber = "BNL-28823",
    doi = "10.1007/BF02902136",
    journal = "Nuovo Cim. A",
    volume = "65",
    pages = "222",
    year = "1981"
}

@article{Du:1993qe,
    author = "Du, Dong-sheng and Xing, Zhi-zhong",
    title = "{Effect of hairpin diagram on two-body nonleptonic B decays and CP violation}",
    eprint = "hep-ph/9307218",
    archivePrefix = "arXiv",
    reportNumber = "BIHEP-TH-93-25",
    doi = "10.1016/0370-2693(93)90510-O",
    journal = "Phys. Lett. B",
    volume = "312",
    pages = "199--204",
    year = "1993"
}

@article{Dery:2019ysp,
    author = "Dery, Avital and Nir, Yosef",
    title = "{Implications of the LHCb discovery of CP violation in charm decays}",
    eprint = "1909.11242",
    archivePrefix = "arXiv",
    primaryClass = "hep-ph",
    doi = "10.1007/JHEP12(2019)104",
    journal = "JHEP",
    volume = "12",
    pages = "104",
    year = "2019"
}

@article{Binosi:2008ig,
    author = "Binosi, D. and Collins, J. and Kaufhold, C. and Theussl, L.",
    title = "{JaxoDraw: A Graphical user interface for drawing Feynman diagrams. Version 2.0 release notes}",
    eprint = "0811.4113",
    archivePrefix = "arXiv",
    primaryClass = "hep-ph",
    reportNumber = "ECT*-08-10",
    doi = "10.1016/j.cpc.2009.02.020",
    journal = "Comput. Phys. Commun.",
    volume = "180",
    pages = "1709--1715",
    year = "2009"
}

@article{Binosi:2003yf,
    author = "Binosi, D. and Theu{\ss}l, L.",
    title = "{JaxoDraw: A Graphical user interface for drawing Feynman diagrams}",
    eprint = "hep-ph/0309015",
    archivePrefix = "arXiv",
    reportNumber = "FTUV-03-0902",
    doi = "10.1016/j.cpc.2004.05.001",
    journal = "Comput. Phys. Commun.",
    volume = "161",
    pages = "76--86",
    year = "2004"
}

@article{Buras:1998ra,
    author = "Buras, Andrzej J. and Silvestrini, Luca",
    title = "{Nonleptonic two-body B decays beyond factorization}",
    eprint = "hep-ph/9812392",
    archivePrefix = "arXiv",
    reportNumber = "TUM-HEP-339-98",
    doi = "10.1016/S0550-3213(99)00712-9",
    journal = "Nucl. Phys. B",
    volume = "569",
    pages = "3--52",
    year = "2000"
}

@article{Iguro:2024uuw,
    author = {Iguro, Syuhei and Nierste, Ulrich and Overduin, Emil and Sch{\"u}{\ss}ler, Maurice},
    title = "{SU(3)$_F$ sum rules for CP asymmetries of $D_{(s)}$ decays}",
    eprint = "2408.03227",
    archivePrefix = "arXiv",
    primaryClass = "hep-ph",
    reportNumber = "KEK-TH-2643, TTP24-029, KA-TP-15-2024",
    doi = "10.1103/PhysRevD.111.035023",
    journal = "Phys. Rev. D",
    volume = "111",
    number = "3",
    pages = "035023",
    year = "2025"
}

@article{Grossman:2006jg,
    author = "Grossman, Yuval and Kagan, Alexander L. and Nir, Yosef",
    title = "{New physics and CP violation in singly Cabibbo suppressed D decays}",
    eprint = "hep-ph/0609178",
    archivePrefix = "arXiv",
    doi = "10.1103/PhysRevD.75.036008",
    journal = "Phys. Rev. D",
    volume = "75",
    pages = "036008",
    year = "2007"
}

@article{Bhattacharya:2012ah,
    author = "Bhattacharya, Bhubanjyoti and Gronau, Michael and Rosner, Jonathan L.",
    title = "{CP asymmetries in singly-Cabibbo-suppressed $D$ decays to two pseudoscalar mesons}",
    eprint = "1201.2351",
    archivePrefix = "arXiv",
    primaryClass = "hep-ph",
    reportNumber = "UDEM-GPP-TH-12-205, TECHNION-PH-12-1, EFI-12-1",
    doi = "10.1103/PhysRevD.85.054014",
    journal = "Phys. Rev. D",
    volume = "85",
    pages = "054014",
    year = "2012"
}

@article{Bhattacharya:2021ndt,
    author = "Bhattacharya, Bhubanjyoti and Datta, Alakabha and Petrov, Alexey A. and Waite, John",
    title = "{Flavor SU(3) in Cabibbo-favored D-meson decays}",
    eprint = "2107.13564",
    archivePrefix = "arXiv",
    primaryClass = "hep-ph",
    reportNumber = "WSU-HEP-2102",
    doi = "10.1007/JHEP10(2021)024",
    journal = "JHEP",
    volume = "10",
    pages = "024",
    year = "2021"
}

@article{Hansen:2012tf,
    author = "Hansen, Maxwell T. and Sharpe, Stephen R.",
    title = "{Multiple-channel generalization of Lellouch-Luscher formula}",
    eprint = "1204.0826",
    archivePrefix = "arXiv",
    primaryClass = "hep-lat",
    doi = "10.1103/PhysRevD.86.016007",
    journal = "Phys. Rev. D",
    volume = "86",
    pages = "016007",
    year = "2012"
}

@article{DiCarlo:2025mnm,
    author = "Di Carlo, Matteo and Erben, Felix and Hansen, Maxwell T.",
    title = "{Long distance contributions to neutral D-meson mixing from lattice QCD}",
    eprint = "2504.16189",
    archivePrefix = "arXiv",
    primaryClass = "hep-lat",
    reportNumber = "CERN-TH-2025-075",
    doi = "10.1007/JHEP07(2025)229",
    journal = "JHEP",
    volume = "07",
    pages = "229",
    year = "2025"
}

@inbook{Zweig:1964jf,
    author = "Zweig, G.",
    editor = "Lichtenberg, D. B. and Rosen, Simon Peter",
    title = "{An SU(3) model for strong interaction symmetry and its breaking. Version 2}",
    booktitle = "{DEVELOPMENTS IN THE QUARK THEORY OF HADRONS. VOL. 1. 1964 - 1978}",
    reportNumber = "CERN-TH-412, NP-14146, PRINT-64-170",
    doi = "10.17181/CERN-TH-412",
    pages = "22--101",
    month = "2",
    year = "1964"
}

@article{Iizuka:1966fk,
    author = "Iizuka, Jugoro",
    title = "{Systematics and phenomenology of meson family}",
    doi = "10.1143/PTPS.37.21",
    journal = "Prog. Theor. Phys. Suppl.",
    volume = "37",
    pages = "21--34",
    year = "1966"
}

@article{Okubo:1963fa,
    author = "Okubo, S.",
    title = "{Phi meson and unitary symmetry model}",
    doi = "10.1016/S0375-9601(63)92548-9",
    journal = "Phys. Lett.",
    volume = "5",
    pages = "165--168",
    year = "1963"
}

@article{LHCb:2025ezf,
    author = "Aaij, Roel and others",
    collaboration = "LHCb",
    title = "{Measurement of $C\!P$ asymmetry in $D^0 \to K^0_{\rm S} K^0_{\rm S}$ decays with the LHCb Upgrade I detector}",
    eprint = "2510.14732",
    archivePrefix = "arXiv",
    primaryClass = "hep-ex",
    reportNumber = "LHCb-PAPER-2025-036, CERN-EP-2025-221",
    month = "10",
    year = "2025"
}

@article{HeavyFlavorAveragingGroupHFLAV:2024ctg,
    author = "Banerjee, Swagato and others",
    collaboration = "Heavy Flavor Averaging Group (HFLAV)",
    title = "{Averages of $b$-hadron, $c$-hadron, and $\tau$-lepton properties as of 2023}",
    eprint = "2411.18639",
    archivePrefix = "arXiv",
    primaryClass = "hep-ex",
    month = "11",
    year = "2024"
}

@article{CLEO:2009fiz,
    author = "Mendez, H. and others",
    collaboration = "CLEO",
    title = "{Measurements of D Meson Decays to Two Pseudoscalar Mesons}",
    eprint = "0906.3198",
    archivePrefix = "arXiv",
    primaryClass = "hep-ex",
    reportNumber = "CLNS-09-2054, CLEO-09-07",
    doi = "10.1103/PhysRevD.81.052013",
    journal = "Phys. Rev. D",
    volume = "81",
    pages = "052013",
    year = "2010"
}

@article{Belle:2011tmj,
    author = "Won, E. and others",
    collaboration = "Belle",
    title = "{Observation of $D^+ \rightarrow K^{+} \eta^{(\prime)}$ and Search for CP Violation in $D^+ \rightarrow \pi^+ \eta^{(\prime)}$ Decays}",
    eprint = "1107.0553",
    archivePrefix = "arXiv",
    primaryClass = "hep-ex",
    doi = "10.1103/PhysRevLett.107.221801",
    journal = "Phys. Rev. Lett.",
    volume = "107",
    pages = "221801",
    year = "2011"
}

@article{Muller:2015rna,
    author = {M\"uller, Sarah and Nierste, Ulrich and Schacht, Stefan},
    title = "{Sum Rules of Charm CP Asymmetries beyond the SU(3)$_F$ Limit}",
    eprint = "1506.04121",
    archivePrefix = "arXiv",
    primaryClass = "hep-ph",
    reportNumber = "TTP15-020",
    doi = "10.1103/PhysRevLett.115.251802",
    journal = "Phys. Rev. Lett.",
    volume = "115",
    number = "25",
    pages = "251802",
    year = "2015"
}

@article{Bhattacharya:2009ps,
    author = "Bhattacharya, Bhubanjyoti and Rosner, Jonathan L.",
    title = "{Charmed meson decays to two pseudoscalars}",
    eprint = "0911.2812",
    archivePrefix = "arXiv",
    primaryClass = "hep-ph",
    reportNumber = "EFI-09-32",
    doi = "10.1103/PhysRevD.81.014026",
    journal = "Phys. Rev. D",
    volume = "81",
    pages = "014026",
    year = "2010"
}

@article{Bolognani:2024zno,
    author = "Bolognani, Carolina and Nierste, Ulrich and Schacht, Stefan and Vos, K. Keri",
    title = "{Anatomy of non-leptonic two-body decays of charmed mesons into final states with {\ensuremath{\eta}}'}",
    eprint = "2410.08138",
    archivePrefix = "arXiv",
    primaryClass = "hep-ph",
    reportNumber = "Nikhef 2024-017, TTP24-040",
    doi = "10.1007/JHEP05(2025)148",
    journal = "JHEP",
    volume = "05",
    pages = "148",
    year = "2025"
}

@article{BESIII:2025ykz,
    author = "Ablikim, Medina and others",
    collaboration = "BESIII",
    title = "{Measurements of the absolute branching fractions of the doubly Cabibbo-suppressed decays $D^+\to K^+\pi^0$, $D^+\to K^+\eta$ and $D^+\to K^+\eta^{\prime}$}",
    eprint = "2506.15533",
    archivePrefix = "arXiv",
    primaryClass = "hep-ex",
    month = "6",
    year = "2025"
}

@article{LHCb:2023mwc,
    author = "Aaij, R. and others",
    collaboration = "LHCb",
    title = "{Search for $CP$ violation in the phase space of $D^0 \to \pi^-\pi^+\pi^0$ decays with the energy test}",
    eprint = "2306.12746",
    archivePrefix = "arXiv",
    primaryClass = "hep-ex",
    reportNumber = "CERN-EP-2023-103, LHCb-PAPER-2023-005",
    doi = "10.1007/JHEP09(2023)129",
    journal = "JHEP",
    volume = "09",
    pages = "129",
    year = "2023",
    note = "[Erratum: JHEP 04, 040 (2024)]"
}

@article{LHCb:2023rae,
    author = "Aaij, Roel and others",
    collaboration = "LHCb",
    title = "{Search for CP violation in the phase space of $ {D}^0\to {K}_S^0{K}^{\pm }{\pi}^{\mp } $ decays with the energy test}",
    eprint = "2310.19397",
    archivePrefix = "arXiv",
    primaryClass = "hep-ex",
    reportNumber = "LHCb-PAPER-2023-019, CERN-EP-2023-231",
    doi = "10.1007/JHEP03(2024)107",
    journal = "JHEP",
    volume = "03",
    pages = "107",
    year = "2024"
}

@article{CMS:2024hsv,
    author = "Hayrapetyan, Aram and others",
    collaboration = "CMS",
    title = "{Search for $CP$ violation in D$^0$$\to$ K$^0_\mathrm{S}$K$^0_\mathrm{S}$ decays in proton-proton collisions at $\sqrt{s}$ = 13 TeV}",
    eprint = "2405.11606",
    archivePrefix = "arXiv",
    primaryClass = "hep-ex",
    reportNumber = "CMS-BPH-23-005, CERN-EP-2024-120",
    month = "5",
    year = "2024"
}

@article{Belle-II:2023vra,
    author = "Adachi, I. and others",
    collaboration = "Belle-II",
    title = "{Novel method for the identification of the production flavor of neutral charmed mesons}",
    eprint = "2304.02042",
    archivePrefix = "arXiv",
    primaryClass = "hep-ex",
    reportNumber = "Belle II Preprint 2023-005, KEK Preprint 2023-1",
    doi = "10.1103/PhysRevD.107.112010",
    journal = "Phys. Rev. D",
    volume = "107",
    number = "11",
    pages = "112010",
    year = "2023"
}

@article{Belle-II:2018jsg,
    author = "Altmannshofer, W. and others",
    editor = "Kou, E. and Urquijo, P.",
    collaboration = "Belle-II",
    title = "{The Belle II Physics Book}",
    eprint = "1808.10567",
    archivePrefix = "arXiv",
    primaryClass = "hep-ex",
    reportNumber = "KEK Preprint 2018-27, BELLE2-PUB-PH-2018-001, FERMILAB-PUB-18-398-T, JLAB-THY-18-2780, INT-PUB-18-047, UWThPh 2018-26",
    doi = "10.1093/ptep/ptz106",
    journal = "PTEP",
    volume = "2019",
    number = "12",
    pages = "123C01",
    year = "2019",
    note = "[Erratum: PTEP 2020, 029201 (2020)]"
}

@article{LHCb:2018roe,
    author = "Aaij, Roel and others",
    collaboration = "LHCb",
    title = "{Physics case for an LHCb Upgrade II - Opportunities in flavour physics, and beyond, in the HL-LHC era}",
    eprint = "1808.08865",
    archivePrefix = "arXiv",
    primaryClass = "hep-ex",
    reportNumber = "LHCB-PUB-2018-009, CERN-LHCC-2018-027",
    month = "8",
    year = "2018"
}

@article{BESIII:2022xhe,
    author = "Ablikim, Medina and others",
    collaboration = "BESIII",
    title = "{Measurements of absolute branching fractions of $D^0\to K_L^0\phi$, $K_L^0\eta$, $K_L^0\omega$, and $K_L^0\eta^{\prime}$}",
    eprint = "2202.13601",
    archivePrefix = "arXiv",
    primaryClass = "hep-ex",
    doi = "10.1103/PhysRevD.105.092010",
    journal = "Phys. Rev. D",
    volume = "105",
    number = "9",
    pages = "092010",
    year = "2022"
}

@article{Belle:2021dfa,
    author = "Li, L. K. and others",
    collaboration = "Belle",
    title = "{Measurement of branching fractions and search for $CP$ violation in $D^{0}\to\pi^{+}\pi^{-}\eta$, $D^{0}\to K^{+}K^{-}\eta$, and $D^{0}\to\phi\eta$ at Belle}",
    eprint = "2106.04286",
    archivePrefix = "arXiv",
    primaryClass = "hep-ex",
    reportNumber = "KEK Preprint 2021-8, Belle Preprint 2021-11, UCHEP-21-03",
    doi = "10.1007/JHEP09(2021)075",
    journal = "JHEP",
    volume = "09",
    pages = "075",
    year = "2021"
}

@article{LHCb:2022pxf,
    author = "Aaij, Roel and others",
    collaboration = "LHCb",
    title = "{Measurement of CP asymmetries in $ {D}_{(s)}^{+}\to \eta {\pi}^{+} $ and $ {D}_{(s)}^{+}\to {\eta}^{\prime }{\pi}^{+} $ decays}",
    eprint = "2204.12228",
    archivePrefix = "arXiv",
    primaryClass = "hep-ex",
    reportNumber = "CERN-EP-2022-035, LHCb-PAPER-2021-051",
    doi = "10.1007/JHEP04(2023)081",
    journal = "JHEP",
    volume = "04",
    pages = "081",
    year = "2023"
}

@article{LHCb:2021rou,
    author = "Aaij, Roel and others",
    collaboration = "LHCb",
    title = "{Search for CP violation in $ {D}_{(s)}^{+}\to {h}^{+}{\pi}^0 $ and $ {D}_{(s)}^{+}\to {h}^{+}\eta $ decays}",
    eprint = "2103.11058",
    archivePrefix = "arXiv",
    primaryClass = "hep-ex",
    reportNumber = "LHCb-PAPER-2021-001, CERN-EP-2021-036",
    doi = "10.1007/JHEP06(2021)019",
    journal = "JHEP",
    volume = "06",
    pages = "019",
    year = "2021"
}

@article{Belle:2021ygw,
    author = "Guan, Y. and others",
    collaboration = "Belle",
    title = "{Measurement of branching fractions and $CP$ asymmetries for $D_s^{+} \rightarrow K^+ (\eta, \pi^0) $ and $D_s^{+} \rightarrow \pi^+ (\eta, \pi^0)$ decays at Belle}",
    eprint = "2103.09969",
    archivePrefix = "arXiv",
    primaryClass = "hep-ex",
    reportNumber = "Belle Preprint 2021-04; KEK Preprint 2020-46; UC Preprint
  UCHEP-21-02",
    doi = "10.1103/PhysRevD.103.112005",
    journal = "Phys. Rev. D",
    volume = "103",
    pages = "112005",
    year = "2021"
}

@article{Gavrilova:2022hbx,
    author = "Gavrilova, Margarita and Grossman, Yuval and Schacht, Stefan",
    title = "{The mathematical structure of U-spin amplitude sum rules}",
    eprint = "2205.12975",
    archivePrefix = "arXiv",
    primaryClass = "hep-ph",
    doi = "10.1007/JHEP08(2022)278",
    journal = "JHEP",
    volume = "08",
    pages = "278",
    year = "2022"
}

@article{Gavrilova:2024npn,
    author = "Gavrilova, Margarita and Schacht, Stefan",
    title = "{Systematics of U-Spin Sum Rules for Systems with Direct Sums}",
    eprint = "2409.03830",
    archivePrefix = "arXiv",
    primaryClass = "hep-ph",
    month = "9",
    year = "2024"
}

@article{Brod:2012ud,
    author = "Brod, Joachim and Grossman, Yuval and Kagan, Alexander L. and Zupan, Jure",
    title = "{A Consistent Picture for Large Penguins in $D \rightarrow \pi^+ \pi^-, K^+ K^-$}",
    eprint = "1203.6659",
    archivePrefix = "arXiv",
    primaryClass = "hep-ph",
    reportNumber = "UCHEP-12-02",
    doi = "10.1007/JHEP10(2012)161",
    journal = "JHEP",
    volume = "10",
    pages = "161",
    year = "2012"
}

@article{Buras:1985xv,
    author = "Buras, A. J. and Gerard, J. M. and Ruckl, R.",
    title = "{$1/N$ Expansion for Exclusive and Inclusive Charm Decays}",
    reportNumber = "MPI-PAE-PTH-41-85",
    doi = "10.1016/0550-3213(86)90200-2",
    journal = "Nucl. Phys. B",
    volume = "268",
    pages = "16--48",
    year = "1986"
}

@article{tHooft:1973alw,
    author = "'t Hooft, Gerard",
    editor = "Taylor, J. C.",
    title = "{A Planar Diagram Theory for Strong Interactions}",
    reportNumber = "CERN-TH-1786",
    doi = "10.1016/0550-3213(74)90154-0",
    journal = "Nucl. Phys. B",
    volume = "72",
    pages = "461",
    year = "1974"
}

@article{Golden:1989qx,
    author = "Golden, Mitchell and Grinstein, Benjamin",
    title = "{Enhanced CP Violations in Hadronic Charm Decays}",
    reportNumber = "FERMILAB-PUB-89-048-T",
    doi = "10.1016/0370-2693(89)90353-5",
    journal = "Phys. Lett. B",
    volume = "222",
    pages = "501--506",
    year = "1989"
}

@article{Cheng:2012wr,
    author = "Cheng, Hai-Yang and Chiang, Cheng-Wei",
    title = "{Direct CP violation in two-body hadronic charmed meson decays}",
    eprint = "1201.0785",
    archivePrefix = "arXiv",
    primaryClass = "hep-ph",
    doi = "10.1103/PhysRevD.85.034036",
    journal = "Phys. Rev. D",
    volume = "85",
    pages = "034036",
    year = "2012",
    note = "[Erratum: Phys.Rev.D 85, 079903 (2012)]"
}

@article{Feldmann:2012js,
    author = "Feldmann, Thorsten and Nandi, Soumitra and Soni, Amarjit",
    title = "{Repercussions of Flavour Symmetry Breaking on CP Violation in D-Meson Decays}",
    eprint = "1202.3795",
    archivePrefix = "arXiv",
    primaryClass = "hep-ph",
    reportNumber = "SI-HEP-2012-03",
    doi = "10.1007/JHEP06(2012)007",
    journal = "JHEP",
    volume = "06",
    pages = "007",
    year = "2012"
}

@article{Grossman:2012ry,
    author = "Grossman, Yuval and Robinson, Dean J.",
    title = "{SU(3) Sum Rules for Charm Decay}",
    eprint = "1211.3361",
    archivePrefix = "arXiv",
    primaryClass = "hep-ph",
    doi = "10.1007/JHEP04(2013)067",
    journal = "JHEP",
    volume = "04",
    pages = "067",
    year = "2013"
}

@article{Nierste:2015zra,
    author = "Nierste, Ulrich and Schacht, Stefan",
    title = "{CP Violation in $D^0\rightarrow K_SK_S$}",
    eprint = "1508.00074",
    archivePrefix = "arXiv",
    primaryClass = "hep-ph",
    reportNumber = "TTP15-027",
    doi = "10.1103/PhysRevD.92.054036",
    journal = "Phys. Rev. D",
    volume = "92",
    number = "5",
    pages = "054036",
    year = "2015"
}

@article{Nierste:2017cua,
    author = "Nierste, Ulrich and Schacht, Stefan",
    title = "{Neutral $D\rightarrow K K^*$ decays as discovery channels for charm CP violation}",
    eprint = "1708.03572",
    archivePrefix = "arXiv",
    primaryClass = "hep-ph",
    reportNumber = "TTP17-033",
    doi = "10.1103/PhysRevLett.119.251801",
    journal = "Phys. Rev. Lett.",
    volume = "119",
    number = "25",
    pages = "251801",
    year = "2017"
}

@article{Grossman:2018ptn,
    author = "Grossman, Yuval and Schacht, Stefan",
    title = "{U-Spin Sum Rules for CP Asymmetries of Three-Body Charmed Baryon Decays}",
    eprint = "1811.11188",
    archivePrefix = "arXiv",
    primaryClass = "hep-ph",
    doi = "10.1103/PhysRevD.99.033005",
    journal = "Phys. Rev. D",
    volume = "99",
    number = "3",
    pages = "033005",
    year = "2019"
}

@article{Cheng:2019ggx,
    author = "Cheng, Hai-Yang and Chiang, Cheng-Wei",
    title = "{Revisiting CP violation in $D\to P\!P$ and $V\!P$ decays}",
    eprint = "1909.03063",
    archivePrefix = "arXiv",
    primaryClass = "hep-ph",
    doi = "10.1103/PhysRevD.100.093002",
    journal = "Phys. Rev. D",
    volume = "100",
    number = "9",
    pages = "093002",
    year = "2019"
}

@article{Grossman:2012eb,
    author = "Grossman, Yuval and Kagan, Alexander L. and Zupan, Jure",
    title = "{Testing for new physics in singly Cabibbo suppressed D decays}",
    eprint = "1204.3557",
    archivePrefix = "arXiv",
    primaryClass = "hep-ph",
    reportNumber = "UCHEP-12-03",
    doi = "10.1103/PhysRevD.85.114036",
    journal = "Phys. Rev. D",
    volume = "85",
    pages = "114036",
    year = "2012"
}

@article{Gavrilova:2023fzy,
    author = "Gavrilova, Margarita and Grossman, Yuval and Schacht, Stefan",
    title = "{Determination of the D\textrightarrow{}\ensuremath{\pi}\ensuremath{\pi} ratio of penguin over tree diagrams}",
    eprint = "2312.10140",
    archivePrefix = "arXiv",
    primaryClass = "hep-ph",
    doi = "10.1103/PhysRevD.109.033011",
    journal = "Phys. Rev. D",
    volume = "109",
    number = "3",
    pages = "033011",
    year = "2024"
}

@article{Bause:2022jes,
    author = {Bause, Rigo and Gisbert, Hector and Hiller, Gudrun and H\"ohne, Tim and Litim, Daniel F. and Steudtner, Tom},
    title = "{U-spin-CP anomaly in charm}",
    eprint = "2210.16330",
    archivePrefix = "arXiv",
    primaryClass = "hep-ph",
    reportNumber = "DO-TH 22/25",
    doi = "10.1103/PhysRevD.108.035005",
    journal = "Phys. Rev. D",
    volume = "108",
    number = "3",
    pages = "035005",
    year = "2023"
}

@article{Cerri:2018ypt,
    author = "Cerri, A. and others",
    editor = "Dainese, Andrea and Mangano, Michelangelo and Meyer, Andreas B. and Nisati, Aleandro and Salam, Gavin and Vesterinen, Mika Anton",
    title = "{Report from Working Group 4}: {Opportunities in Flavour Physics at the HL-LHC and HE-LHC}",
    eprint = "1812.07638",
    archivePrefix = "arXiv",
    primaryClass = "hep-ph",
    reportNumber = "CERN-LPCC-2018-06",
    doi = "10.23731/CYRM-2019-007.867",
    journal = "CERN Yellow Rep. Monogr.",
    volume = "7",
    pages = "867--1158",
    year = "2019"
}

@article{Belle:2023bzn,
    author = "Moon, H. K. and others",
    collaboration = "Belle",
    title = "{Search for CP violation in $D_{(s)}^+\rightarrow K^+K_S^0h^+h^-$ (h=K, \ensuremath{\pi}) decays and observation of the Cabibbo-suppressed decay $D_s^+\rightarrow K^+K^-K_S^0\pi^+$}",
    eprint = "2305.11405",
    archivePrefix = "arXiv",
    primaryClass = "hep-ex",
    reportNumber = "Belle Preprint:2023-08, KEK Preprint:2023-4",
    doi = "10.1103/PhysRevD.108.L111102",
    journal = "Phys. Rev. D",
    volume = "108",
    number = "11",
    pages = "L111102",
    year = "2023"
}

@article{ParticleDataGroup:2024cfk,
    author = "Navas, S. and others",
    collaboration = "Particle Data Group",
    title = "{Review of particle physics}",
    doi = "10.1103/PhysRevD.110.030001",
    journal = "Phys. Rev. D",
    volume = "110",
    number = "3",
    pages = "030001",
    year = "2024"
}

@article{Schacht:2022kuj,
    author = "Schacht, Stefan",
    title = "{A U-spin anomaly in charm CP violation}",
    eprint = "2207.08539",
    archivePrefix = "arXiv",
    primaryClass = "hep-ph",
    doi = "10.1007/JHEP03(2023)205",
    journal = "JHEP",
    volume = "03",
    pages = "205",
    year = "2023"
}

@article{Dery:2021mll,
    author = "Dery, Avital and Grossman, Yuval and Schacht, Stefan and Soffer, Abner",
    title = "{Probing the $\Delta U=0$ rule in three body charm decays}",
    eprint = "2101.02560",
    archivePrefix = "arXiv",
    primaryClass = "hep-ph",
    reportNumber = "MAN/HEP/2020/015",
    doi = "10.1007/JHEP05(2021)179",
    journal = "JHEP",
    volume = "05",
    pages = "179",
    year = "2021"
}

@article{Schacht:2021jaz,
    author = "Schacht, Stefan and Soni, Amarjit",
    title = "{Enhancement of charm CP violation due to nearby resonances}",
    eprint = "2110.07619",
    archivePrefix = "arXiv",
    primaryClass = "hep-ph",
    doi = "10.1016/j.physletb.2021.136855",
    journal = "Phys. Lett. B",
    volume = "825",
    pages = "136855",
    year = "2022"
}

@article{Franco:2012ck,
    author = "Franco, Enrico and Mishima, Satoshi and Silvestrini, Luca",
    title = "{The Standard Model confronts CP violation in $D^0 \to \pi^+\pi^-$ and $D^0 \to K^+K^-$}",
    eprint = "1203.3131",
    archivePrefix = "arXiv",
    primaryClass = "hep-ph",
    doi = "10.1007/JHEP05(2012)140",
    journal = "JHEP",
    volume = "05",
    pages = "140",
    year = "2012"
}

@article{Gersabeck:2011xj,
    author = "Gersabeck, M. and Alexander, M. and Borghi, S. and Gligorov, V. V. and Parkes, C.",
    title = "{On the interplay of direct and indirect CP violation in the charm sector}",
    eprint = "1111.6515",
    archivePrefix = "arXiv",
    primaryClass = "hep-ex",
    doi = "10.1088/0954-3899/39/4/045005",
    journal = "J. Phys. G",
    volume = "39",
    pages = "045005",
    year = "2012"
}

@article{LHCb:2022lry,
    author = "Aaij, R. and others",
    collaboration = "LHCb",
    title = "{Measurement of the Time-Integrated CP Asymmetry in $D^0\rightarrow K^-K^+$ Decays}",
    eprint = "2209.03179",
    archivePrefix = "arXiv",
    primaryClass = "hep-ex",
    reportNumber = "CERN-EP-2022-163, LHCb-PAPER-2022-024",
    doi = "10.1103/PhysRevLett.131.091802",
    journal = "Phys. Rev. Lett.",
    volume = "131",
    number = "9",
    pages = "091802",
    year = "2023"
}

@article{Kagan:2020vri,
    author = "Kagan, Alexander L. and Silvestrini, Luca",
    title = "{Dispersive and absorptive $CP$ violation in $D^0- \overline{D^0}$ mixing}",
    eprint = "2001.07207",
    archivePrefix = "arXiv",
    primaryClass = "hep-ph",
    reportNumber = "CERN-TH-2020-011",
    doi = "10.1103/PhysRevD.103.053008",
    journal = "Phys. Rev. D",
    volume = "103",
    number = "5",
    pages = "053008",
    year = "2021"
}

@article{Pirtskhalava:2011va,
    author = "Pirtskhalava, David and Uttayarat, Patipan",
    title = "{CP Violation and Flavor SU(3) Breaking in D-meson Decays}",
    eprint = "1112.5451",
    archivePrefix = "arXiv",
    primaryClass = "hep-ph",
    reportNumber = "UCSD-PTH-11-21",
    doi = "10.1016/j.physletb.2012.04.039",
    journal = "Phys. Lett. B",
    volume = "712",
    pages = "81--86",
    year = "2012"
}

@article{Grossman:2013lya,
    author = "Grossman, Yuval and Ligeti, Zoltan and Robinson, Dean J.",
    title = "{More Flavor SU(3) Tests for New Physics in CP Violating B Decays}",
    eprint = "1308.4143",
    archivePrefix = "arXiv",
    primaryClass = "hep-ph",
    doi = "10.1007/JHEP01(2014)066",
    journal = "JHEP",
    volume = "01",
    pages = "066",
    year = "2014"
}

@article{Muller:2015lua,
      author         = "{M\"uller}, Sarah and Nierste, Ulrich and Schacht, Stefan",
      title          = "{Topological amplitudes in $D$ decays to two
                        pseudoscalars: A global analysis with linear $SU(3)_F$
                        breaking}",
      journal        = "Phys. Rev.",
      volume         = "D92",
      year           = "2015",
      number         = "1",
      pages          = "014004",
      doi            = "10.1103/PhysRevD.92.014004",
      eprint         = "1503.06759",
      archivePrefix  = "arXiv",
      primaryClass   = "hep-ph",
      reportNumber   = "TTP15-015",
      SLACcitation   = "%%CITATION = ARXIV:1503.06759;%%"
}

@article{Hiller:2012xm,
    author = "Hiller, Gudrun and Jung, Martin and Schacht, Stefan",
    title = "{SU(3)-flavor anatomy of nonleptonic charm decays}",
    eprint = "1211.3734",
    archivePrefix = "arXiv",
    primaryClass = "hep-ph",
    reportNumber = "DO-TH-12-22",
    doi = "10.1103/PhysRevD.87.014024",
    journal = "Phys. Rev. D",
    volume = "87",
    number = "1",
    pages = "014024",
    year = "2013"
}

@article{Khodjamirian:2017zdu,
    author = "Khodjamirian, Alexander and Petrov, Alexey A.",
    title = "{Direct CP asymmetry in $D\to \pi^-\pi^+$ and $D\to K^-K^+$ in QCD-based approach}",
    eprint = "1706.07780",
    archivePrefix = "arXiv",
    primaryClass = "hep-ph",
    reportNumber = "SI-HEP-2017-12, QFET-2017-09, WSU-HEP-1709",
    doi = "10.1016/j.physletb.2017.09.070",
    journal = "Phys. Lett. B",
    volume = "774",
    pages = "235--242",
    year = "2017"
}

@article{LHCb:2019hro,
    author = "Aaij, Roel and others",
    collaboration = "LHCb",
    title = "{Observation of CP Violation in Charm Decays}",
    eprint = "1903.08726",
    archivePrefix = "arXiv",
    primaryClass = "hep-ex",
    reportNumber = "LHCb-PAPER-2019-006, CERN-EP-2019-042",
    doi = "10.1103/PhysRevLett.122.211803",
    journal = "Phys. Rev. Lett.",
    volume = "122",
    number = "21",
    pages = "211803",
    year = "2019"
}

@article{Chala:2019fdb,
    author = "Chala, Mikael and Lenz, Alexander and Rusov, Aleksey V. and Scholtz, Jakub",
    title = "{$\Delta A_{CP}$ within the Standard Model and beyond}",
    eprint = "1903.10490",
    archivePrefix = "arXiv",
    primaryClass = "hep-ph",
    reportNumber = "IPPP/19/25",
    doi = "10.1007/JHEP07(2019)161",
    journal = "JHEP",
    volume = "07",
    pages = "161",
    year = "2019"
}

@book{kraft1988software,
  title={A Software Package for Sequential Quadratic Programming},
  author={Kraft, D.},
  series={Deutsche Forschungs- und Versuchsanstalt f{\"u}r Luft- und Raumfahrt K{\"o}ln: Forschungsbericht},
  url={https://books.google.nl/books?id=4rKaGwAACAAJ},
  year={1988},
  publisher={Wiss. Berichtswesen d. DFVLR}
}

@ARTICLE{2020SciPy-NMeth,
  author  = {Virtanen, Pauli and Gommers, Ralf and Oliphant, Travis E. and
            Haberland, Matt and Reddy, Tyler and Cournapeau, David and
            Burovski, Evgeni and Peterson, Pearu and Weckesser, Warren and
            Bright, Jonathan and {van der Walt}, St{\'e}fan J. and
            Brett, Matthew and Wilson, Joshua and Millman, K. Jarrod and
            Mayorov, Nikolay and Nelson, Andrew R. J. and Jones, Eric and
            Kern, Robert and Larson, Eric and Carey, C J and
            Polat, {\.I}lhan and Feng, Yu and Moore, Eric W. and
            {VanderPlas}, Jake and Laxalde, Denis and Perktold, Josef and
            Cimrman, Robert and Henriksen, Ian and Quintero, E. A. and
            Harris, Charles R. and Archibald, Anne M. and
            Ribeiro, Ant{\^o}nio H. and Pedregosa, Fabian and
            {van Mulbregt}, Paul and {SciPy 1.0 Contributors}},
  title   = {{{SciPy} 1.0: Fundamental Algorithms for Scientific
            Computing in Python}},
  journal = {Nature Methods},
  year    = {2020},
  volume  = {17},
  pages   = {261--272},
  adsurl  = {https://rdcu.be/b08Wh},
  doi     = {10.1038/s41592-019-0686-2},
}

@article{LHCb:2024rkp,
    author = "Aaij, Roel and others",
    collaboration = "LHCb",
    title = "{Measurement of CP Violation Observables in $D^+\rightarrow K^-K^+\pi^+$ Decays}",
    eprint = "2409.01414",
    archivePrefix = "arXiv",
    primaryClass = "hep-ex",
    reportNumber = "CERN-EP-2024-204, LHCb-PAPER-2024-019",
    doi = "10.1103/PhysRevLett.133.251801",
    journal = "Phys. Rev. Lett.",
    volume = "133",
    number = "25",
    pages = "251801",
    year = "2024"
}

@article{Lenz:2023rlq,
    author = "Lenz, Alexander and Piscopo, Maria Laura and Rusov, Aleksey V.",
    title = "{Two body non-leptonic D$^{0}$ decays from LCSR and implications for ${\Delta a}_{{\text{CP}}}^{{\text{dir}}}$}",
    eprint = "2312.13245",
    archivePrefix = "arXiv",
    primaryClass = "hep-ph",
    reportNumber = "SI-HEP-2023-34, P3H-23-105",
    doi = "10.1007/JHEP03(2024)151",
    journal = "JHEP",
    volume = "03",
    pages = "151",
    year = "2024"
}

@article{Grossman:2019xcj,
    author = "Grossman, Yuval and Schacht, Stefan",
    title = "{The emergence of the $\Delta U=0$ rule in charm physics}",
    eprint = "1903.10952",
    archivePrefix = "arXiv",
    primaryClass = "hep-ph",
    doi = "10.1007/JHEP07(2019)020",
    journal = "JHEP",
    volume = "07",
    pages = "020",
    year = "2019"
}

@article{Brod:2011re,
    author = "Brod, Joachim and Kagan, Alexander L. and Zupan, Jure",
    title = "{Size of direct CP violation in singly Cabibbo-suppressed D decays}",
    eprint = "1111.5000",
    archivePrefix = "arXiv",
    primaryClass = "hep-ph",
    reportNumber = "UCHEP-11-13",
    doi = "10.1103/PhysRevD.86.014023",
    journal = "Phys. Rev. D",
    volume = "86",
    pages = "014023",
    year = "2012"
}

@article{Altmannshofer:2012ur,
    author = "Altmannshofer, Wolfgang and Primulando, Reinard and Yu, Chiu-Tien and Yu, Felix",
    title = "{New Physics Models of Direct CP Violation in Charm Decays}",
    eprint = "1202.2866",
    archivePrefix = "arXiv",
    primaryClass = "hep-ph",
    reportNumber = "FERMILAB-PUB-12-034-T",
    doi = "10.1007/JHEP04(2012)049",
    journal = "JHEP",
    volume = "04",
    pages = "049",
    year = "2012"
}

@article{Bediaga:2022sxw,
    author = "Bediaga, Ignacio and Frederico, Tobias and Magalh{\~a}es, Patricia C.",
    title = "{Enhanced Charm CP Asymmetries from Final State Interactions}",
    eprint = "2203.04056",
    archivePrefix = "arXiv",
    primaryClass = "hep-ph",
    doi = "10.1103/PhysRevLett.131.051802",
    journal = "Phys. Rev. Lett.",
    volume = "131",
    number = "5",
    pages = "051802",
    year = "2023"
}

@article{Pich:2023kim,
    author = "Pich, Antonio and Solomonidi, Eleftheria and Vale Silva, Luiz",
    title = "{Final-state interactions in the CP asymmetries of charm-meson two-body decays}",
    eprint = "2305.11951",
    archivePrefix = "arXiv",
    primaryClass = "hep-ph",
    doi = "10.1103/PhysRevD.108.036026",
    journal = "Phys. Rev. D",
    volume = "108",
    number = "3",
    pages = "036026",
    year = "2023"
}

@article{LHCb:2023qne,
    author = "Aaij, R. and others",
    collaboration = "LHCb",
    title = "{Search for $\it{CP}$ violation in $D_{(s)}^{+}\rightarrow K^{-}K^{+}K^{+}$ decays}",
    eprint = "2303.04062",
    archivePrefix = "arXiv",
    primaryClass = "hep-ex",
    reportNumber = "LHCb-PAPER-2022-042, CERN-EP-2023-010",
    doi = "10.1007/JHEP07(2023)067",
    journal = "JHEP",
    volume = "07",
    pages = "067",
    year = "2023"
}

\end{document}